\RequirePackage{fix-cm}

\documentclass[twocolumn]{svjour3}          
\usepackage{lipsum}
\smartqed   
\usepackage{graphicx}
\usepackage{epstopdf}
\usepackage{amsmath}
\begin{document} \sloppy
\title{A Finite-Element Coarse-Grid Projection Method: A Dual Acceleration/Mesh Refinement Tool for Incompressible Flows
}



\author{A. Kashefi         \and
        A. E. Staples 
}


\institute{A. Kashefi \at
              Engineering Science and Mechanics Program, Virginia Tech, Blacksburg, VA 24061, USA \\
              \email{kashefi@vt.edu}             \\
           A. E. Staples \at
              Faculty of Engineering, Engineering Science and Mechanics Program, Virginia Tech, Blacksburg, VA 24061, USA \\
              \email{aestaples@vt.edu}           
}

\date{Received: date / Accepted: date}

\maketitle

\begin{abstract}
Coarse grid projection (CGP) methodology is a novel multigrid method for systems involving decoupled nonlinear evolution equations and linear elliptic Poisson equations. The nonlinear equations are solved on a fine grid and the linear equations are solved on a corresponding coarsened grid. Mapping operators execute data transfer between the grids. The CGP framework is constructed upon spatial and temporal discretization schemes. This framework has been established for finite volume/difference discretizations as well as explicit time integration methods. In this article we present for the first time a version of CGP for finite element discretizations, which uses a semi-implicit time integration scheme. The mapping functions correspond to the finite-element shape functions. With the novel data structure introduced, the mapping computational cost becomes insignificant. We apply CGP to pressure-correction schemes used for the incompressible Navier-Stokes flow computations. This version is validated on standard test cases with realistic boundary conditions using unstructured triangular meshes. We also pioneer investigations of the effects of CGP on the accuracy of the pressure field. It is found that although CGP reduces the pressure field accuracy, it preserves the accuracy of the pressure gradient and thus the velocity field, while achieving speedup factors ranging from approximately 2 to 30. Exploring the influence of boundary conditions on CGP, the minimum speedup occurs for velocity Dirichlet boundary conditions, while the maximum speedup occurs for open boundary conditions. We discuss the CGP method as a guide for partial mesh refinement of incompressible flow computations and show its application for simulations of flow over a backward-facing step and flow past a cylinder.
\end{abstract}

\section{Introduction and motivation}
\label{intro}

Since the late 1960s, projection methods [1--3] have been widely used for the numerical simulations of transient incompressible Navier-Stokes equations in extensive scientific areas [4, 5] and industrial fields [6, 7]. The high popularity of these methods is due to the scheme’s abilities to execute the incompressibility constrains using a set of two decoupled elliptic equations: a nonlinear advection-diffusion equation, and subsequently, a linear pressure Poisson equation. Although each of the cascade equations imposes computational costs to the system, Poisson’s equation tends to be the most time-consuming component of the flow simulation in complex geometries [8] as well as in computations with open boundary conditions [6, 9].

To substantially lessen the computational expenses, a common approach is to design robust multigrid (MG) solvers, which are solely devoted to one of the existing elliptic systems (see e.g., Refs. [10--12] for MG solvers for the advection-diffusion equation, and Refs. [13, 14] for those associated with Poisson’s equation). In contrast with the methods cited above, a novel family of multigrid procedures, the so-called Coarse Grid Projection (CGP) methodology by the authors [15--18], involves both the nonlinear and linear equations to efficiently accelerate the computations. In the CGP methodology, the nonlinear momentum is balanced on a fine grid, and the linear Poisson's equation is performed on a corresponding coarsened grid. Mapping functions carry out data transfer between the velocity and pressure fields. In this sense, the CGP methods not only effectively relieve the stiff behavior of the primary Poisson problem, but can also take advantage of any desired fast elliptic solver to achieve large speedups while maintaining excellent to reasonable accuracy. The CGP methodology is provided in detail in Sect. 2.2.

The CGP framework was originally proposed by Lentine et al. in 2010 [15] for visual simulations of inviscid flows for video game applications. The uniform grid finite-volume and explicit forward Euler methods were chosen, respectively, for spatial and temporal discretizations in their numerical simulations. This gridding format led to dealing with a volume-weighted Poisson equation [19] in the presence of a solid object (e.g., a sphere) inside the flow field. Aside from the added complication, the grid format would have caused considerable reductions in visual fidelities, if the authors had not employed the complex mapping operators that they did. This makes the CGP technique less practical for curved boundaries. Furthermore, the explicit time integration restricted their algorithm to low CFL (Courant-Friedrichs-Lewy) numbers, and therefore long run times. In 2014 Jin and Chen [18] implemented the same CGP scheme [15] in the fast fluid dynamics (FFD) models to calibrate the effect of this computational accelerator tool on simulating building airflows. A decrease in spurious fluctuations of ventilation rates was observed, but the maximum achieved speedup was only a factor of approximately 1.50.

In 2013 San and Staples [16] expanded the CGP technique (labeled ``CGPRK3'') to the vorticity-stream function formulation of the incompressible Navier-Stokes equations and demonstrated speedup factors ranging from approximately 2 to 42 in their numerical studies. Additionally, they extended the method in order to dramatically lower computational costs associated with elliptic equations of potential vorticity in quasigeostrophic ocean models [17]. Notwithstanding these successes, the CGPRK3 strategy has four main shortcomings. First, the nine-term full weighting restriction [20] and bilinear prolongation [20] operators used can be exclusively utilized in equally spaced grids. Consequently, one must analytically reformulate all steps according to generalized curvilinear coordinates except in uniform rectangular domains. Second, CGPRK3 applies the third-order Runge-Kutta [21] temporal integration to the non-incremental pressure correction method [1], while the splitting error of this specific projection scheme is irreducibly first-order in time and higher-order temporal integration methods do not improve the overall accuracy [1]. Third, the third-order time stepping scheme unnecessarily forces the CGPRK3 scheme to run the Poisson solver three times at each time step, whereas the primary goal of CGP is to reduce the computational effort that arises from the Poisson equation. Fourth, their suggested mapping procedure becomes more costly than that of the Poisson equation solver for high coarsening levels.

To obviate the aforementioned problems, a semi-Implicit-time-integration Finite-Element version of the CGP method (IFE-CGP) is presented in the current study. The incorporation of a semi-implicit backward time integration results in a simple five-step CGP algorithm with nearly zero cost for the data restriction /prolongation operators. It typically enables flow simulations at large time steps and thus improves speedup [22]. The triangular finite element meshes improve the CGP method in the following ways. First, they enhance the scheme’s capacity for the solution of fluid problems defined on complicated domains, where irregular grids and realistic boundary conditions are unavoidable. Second, they facilitate the design of the required mapping modules so that the restriction/prolongation operators can be optionally equivalent to the shape functions approximating multilevel nested spaces of velocity and pressure fields. Third, the generation of the Laplacian and divergence operators on a coarsened grid is expedited by means of available geometric/algebraic multigrid (GMG/AMG) tools for the finite element method (see e.g., [23--26]). This feature is particularly important for obtaining a sufficiently accurate solution of Poisson's equation in modeling flow over obstacles.

This article's objective is to present a simple, elegant version of the CGP method for finite element discretizations of incompressible fluid flow problems. As accelerating incompressible flow computations is the major application of CGP, speedup rates of the computations and the corresponding reduction in the accuracy of velocity and pressure fields are calculated. Besides this main application, we explore mesh refinement usages of the CGP method for the first time. As a next concern, because the accuracy of the pressure field in projection methods suffers from the negative influence of artificial homogeneous Dirichlet and Neumann boundary conditions on formations of boundary layers [6, 9], the possibility of thickening the layers by the CGP prolongation operator is investigated. Lastly, because the CGP procedure reduces the degree of freedom for the pressure component, a greater reduction in the integrity of the pressure field appears in comparison with the velocity field [16]. On the other hand, the pressure gradient (instead of simply pressure) is applied to a velocity correction step of pressure projection schemes [27]. With these two hypotheses in mind, we examine the effect of the CGP process on variations of both the pressure and its gradient. All the above mentioned numerical challenges are investigated through several representative benchmark problems: the Taylor-Green vortex in a non-trivial geometry, flow over a backward-facing step, and finally flow past a circular cylinder.

This article structured as follows. Sect. 2.1 provides the governing equations for incompressible viscous flows and their finite element formulations. Details on the proposed CGP algorithm are presented in Sect. 2.2. The computational implementation and a computational cost analysis are described, respectively, in Sect. 2.3 and 2.4. Numerical results and their interpretations are collected in Sect. 3. Conclusions and notes for extensions of the work are given in Sect. 4.

\section{Problem formulation}
\subsection{Governing equations}
\label{sec:2}
Mass and momentum conservation of an incompressible isothermal flow of a Newtonian fluid are governed by the Navier-Stokes and continuity equations, with boundary conditions
\begin{equation}
\rho \bigg[\frac{\partial \textbf{\textit{u}}}{\partial t}+(\textbf{\textit{u}}\cdot \nabla)\textbf{\textit{u}}\bigg] -\mu \Delta \textbf{\textit{u}} + \nabla p=\textbf{\textit{f}} \textrm{ in } V,
\end{equation}
\begin{equation}
\nabla \cdot \textbf{\textit{u}}=0 \textrm{ in } V,
\end{equation}
\begin{equation}
\textbf{\textit{u}}=\textbf{\textit{u}}_{\Gamma_D} \textrm{ on } \Gamma_D,
\end{equation}
\begin{equation}
-p\textbf{\textit{n}}+\mu \nabla \textbf{\textit{u}} \cdot \textbf{\textit{n}}=\textbf{\textit{t}}_{\Gamma_N} \textrm{ on } \Gamma_N,
\end{equation}
where $\textbf{\textit{u}}$ and $p$ respectively denote the velocity vector and the absolute pressure in the fluid domain $V$. The external force and stress vectors are represented by $\textbf{\textit{f}}$  and $\textbf{\textit{t}}_{\Gamma_N}$, respectively. $\rho$ is the fluid density and $\mu$ is the dynamic viscosity. The boundary $\Gamma$ of the domain $V$ consists of two non-overlapping subsets of Dirichlet $\Gamma_D$ and Neumann $\Gamma_N$ boundaries, where $\textbf{\textit{n}}$ indicates the outward unit vector normal to them.
The system of equations is temporally integrated by the semi-implicit first-order backward differentiation formula [28] with time increment $\delta t$, and takes the form:
\begin{equation}
\begin{aligned}
\rho \bigg[\frac{\textbf{\textit{u}}^{n+1} - \textbf{\textit{u}}^n}{\delta t}+(\textbf{\textit{u}}^n\cdot \nabla)\textbf{\textit{u}}^{n+1}\bigg] -\mu \Delta  \textbf{\textit{u}}^{n+1} + \nabla p^{n+1} \\
=\textbf{\textit{f}}^{n+1} \textrm{ in } V,
\end{aligned}
\end{equation}

\begin{equation}
\nabla \cdot \textbf{\textit{u}}^{n+1}=0 \textrm{ in } V,
\end{equation}
\begin{equation}
\textbf{\textit{u}}^{n+1}=\textbf{\textit{u}}_{\Gamma_D}^{n+1} \textrm{ on } \Gamma_D,
\end{equation}
\begin{equation}
-p^{n+1}\textbf{\textit{n}}+\mu \nabla \textbf{\textit{u}}^{n+1} \cdot \textbf{\textit{n}}=\textbf{\textit{t}}_{\Gamma_N}^{n+1} \textrm{ on } \Gamma_N.
\end{equation}
In the next stage, the non-incremental pressure-correction method [1] decouples the solution of the velocity and pressure variables. Based on it, at each time step $t^{n+1}$, the output of the momentum equation is a non-divergent vector field called the intermediate velocity $\tilde{\textbf{\textit{u}}}^{n+1}$. Next, the divergence of the intermediate velocity is fed to the source term of the pressure Poisson equation. Finally, the intermediate velocity $\tilde{\textbf{\textit{u}}}^{n+1}$ is corrected using the obtained pressure $p^{n+1}$ such that the end-of-step velocity vector $\textbf{\textit{u}}^{n+1}$ satisfies the incompressibility constraint. The procedure yields two elliptic problems along with one correction equation, expressed as

\begin{equation}
\rho \bigg[\frac{\textbf{\textit{\~u}}^{n+1} - \textbf{\textit{u}}^n}{\delta t}+(\textbf{\textit{u}}^n\cdot \nabla)\textbf{\textit{\~u}}^{n+1}\bigg] -\mu \Delta \textbf{\textit{\~u}}^{n+1}=\textbf{\textit{f}}^{n+1} \textrm{ in } V,
\end{equation}

\begin{equation}
\textbf{\textit{\~u}}^{n+1}=\textbf{\textit{u}}_{\Gamma_D}^{n+1} \textrm{ on } \Gamma_D,
\end{equation}

\begin{equation}
\mu \nabla \textbf{\textit{\~u}}^{n+1} \cdot \textbf{\textit{n}}=\textbf{\textit{t}}_{\Gamma_N}^{n+1} \textrm{ on } \Gamma_N,
\end{equation}

\begin{equation}
\Delta p^{n+1}=\frac{\rho}{\delta t} \nabla \cdot \textbf{\textit{\~u}}^{n+1} \textrm{ in } V,
\end{equation}

\begin{equation}
\nabla p^{n+1} \cdot \textbf{n}=0  \textrm{ on } \Gamma_D,
\end{equation}

\begin{equation}
p^{n+1}=0  \textrm{ on } \Gamma_N,
\end{equation}

\begin{equation}
\textbf{\textit{u}}^{n+1}=\textbf{\textit{\~u}}^{n+1}-\frac{\delta t}{\rho}\nabla p^{n+1}.
\end{equation}
Notice that the Poisson equation is restricted to unrealistic homogenous Neumann boundary conditions, when the velocity is subject to Dirichlet types. Contrarily, natural Neumann conditions for the momentum equation result in boundaries with spurious homogenous Dirichlet pressures.

The velocity and pressure spaces are approximated by the piecewise linear basis functions (\textbf{P}$_1$/\textbf{P}$_1$) of standard Galerkin spectral elements [29]. In this way, the resulting finite element form of Eqs. (9)--(15) is

\begin{equation}
\frac{1}{\delta t}\big(\textbf{M}_v\textrm{\~U}^{n+1}
-\textbf{M}_v\textrm{\~U}^{n}\big)+\big[\textbf{N}+\textbf{L}_v\big]\textrm{\~U}^{n+1}=\textbf{M}_v\textrm{F}^{n+1},
\end{equation}

\begin{equation}
\textbf{L}_p\textrm{P}^{n+1}=\frac{\rho}{\delta t}\textbf{D}\textrm{\~U}^{n+1},
\end{equation}

\begin{equation}
\textbf{M}_v\textrm{U}^{n+1}=\textbf{M}_v\textrm{\~U}^{n+1}-\delta
t\textbf{G}\textrm{P}^{n+1},
\end{equation}
where \textbf{M}$_v$ and \textbf{N} indicate, respectively, the velocity mass and the nonlinear convection matrices. \textbf{L}$_v$, \textbf{L}$_p$, \textbf{D}, and \textbf{G} denote the matrices associated, respectively, to the velocity laplacian, the pressure laplacian, the divergence, and the gradient operators. The vectors {\~U}$^{n+1}$, {U}$^{n+1}$, {F}$^{n+1}$, and {P}$^{n+1}$ contain, respectively, the nodal values of the intermediate velocity, the end-of-step velocity, the forcing term and the pressure at time $t^{n+1}$. The desired boundary conditions are implicitly enforced in the discrete operators. Further details of the elemental matrices are available in references [29, 30].

\textbf{Remark 2.1 }\textit{(On the discrete Brezzi-Babuska condition}[31, 32]\textit{).} Because the projection method overcomes the well-known saddle-point issue of Eqs. (1)--(2), the discrete Brezzi-Babuska condition [31, 32] can be ignored [27] and therefore the degree of the polynomials over the triangular mesh elements is chosen to be equal for the velocity and pressure. In addition, the identical resolutions allow the comparison of computational effectiveness between the current approach and that described in previous works [15--18].

\subsection{Coarse grid projection methodology}
\label{sec:3}

The CGP methodology provides a multiresolution framework for accelerating pressure-correction technique computations by performing part of the computations on a coarsened grid. Alternatively, CGP can be thought as a guide to mesh refinement for pressure-correction schemes. Figure 1 gives a schematic illustration of the IFE-CGP algorithm for triangular finite element meshes. As shown in Fig. 1, at each time step $t^{n+1}$, the intermediate velocity field data \~{U}$_f^{n+1}$ obtained on a fine grid is restricted to a coarsened grid. The divergence of  \~{U}$_c^{n+1}$, the intermediate velocity field data on the coarsened grid, plays the role of the source term in solving the Poisson equation on the coarse grid to detrimine the pressure P$_c^{n+1}$ on the coarse grid; then, the resulting pressure data P$_c^{n+1}$ is prolonged to the fine grid and becomes of P$_f^{n+1}$.

\begin{figure*}
\centering
\includegraphics[width=187 mm]{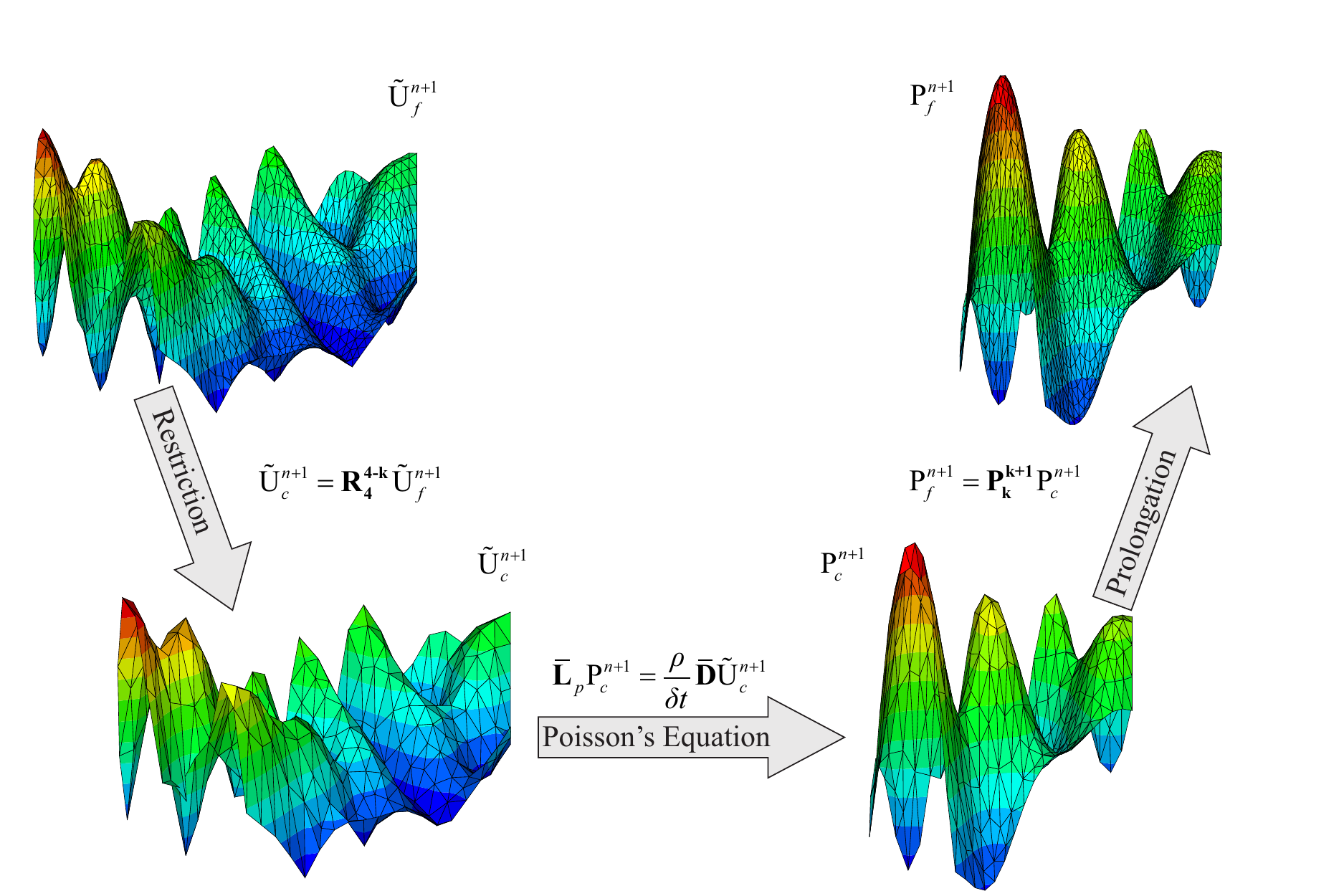}
\caption{Scheme of coarse grid projection methodology involving the restriction and prolongation of the intermediate velocity and pressure data.}
\label{fig:1} 
\end{figure*}

Either GMG or AMG techniques are key numerical tools to conveniently formulate the grid transfer, laplacian, and divergence operators in the IFE-CGP algorithm. In the present work, GMG methods are used, because using AMG methods in the derivation of the divergence operator is challenging for anisotropic triangular grids [33]. Here we generate hierarchical meshes in which each triangular element of a coarse mesh is conformingly subdivided into four new triangles.  As a consequence, if the available finest mesh (with $M$-element resolution) is generated after $l$ refinement levels, a relatively coarse mesh (with a resolution of $N=4^{-l}M$ elements) and its basis functions are available. Generally, one can obtain the corresponding coarsened grid using the space decomposition algorithm discussed for nonuniform triangular grids in the literature (see e.g., Refs. [34, 35]).

In principle, there can be an arbitrary number, $j$, of nested spaces of progressively coarsened grids such that: $V_1 \subset V_2 \subset \cdots \subset V_j=V$. In practice, however, reasonable levels of accuracy can only be obtained for a maximum of three levels of coarsening. We restrict our attention to such cases. Consider a sequence of four nested spaces, $V_1 \subset V_2 \subset V_3 \subset V_4=V$, wherein if $V_{k+1}$ ($k=1, 2, 3$) corresponds to a fine grid, $V_k$ characterizes the space of the next coarsest grid. Recall that for any arbitrary $\textbf{P}_1$ element on $V_k$, there is a piecewise linear shape function $\psi^k$ that is capable of estimating the space of the four sub-elements over $V_{k+1}$. With this in mind, linear interpolation is implemented to construct the prolongation operator $P:V_k \rightarrow V_{k+1}$ and its matrix representation \textbf{P}$_{\textbf{k}}^{\textbf{k+1}}$. Mathematically, the transpose of the prolongation operator is a feasible option for the definition of the restriction operator $R$ [26]. In spite of this fact, the restriction matrix $\textbf{R}_{\textbf{4}}^{\textbf{4-k}}$ is designed so that fine data of $V_4$ is directly injected throughout the coarse grid $V_{4-k}$ only if the data belongs to a common node between the two spaces; otherwise, it cannot be accessed by the coarse grid. Unlike $\textbf{P}_{\textbf{k}}^{\textbf{k+1}}$, $\textbf{R}_{\textbf{4}}^{\textbf{4-k}}$ is able to directly connect two non-nested spaces. It is worth noting that the CGP method is compatible with other advanced data interpolation/extrapolation architectures (see e.g., [36]); however, even the simple mapping techniques introduced here are adequate.

In the GMG method used, a relatively coarse mesh, $V_{4-k}$, with $\{ \psi_1^{4-k}, \psi_2^{4-k}, \cdots, \psi_{\#(V_{4-k})}^{4-k} \}$ as a basis is directly accessible. Hence, the relevant laplacian ($\bar{\textbf{L}}_p$) and divergence ($\bar{\textbf{D}}$) operators are derived by taking the inner products of a coarse-grid finite element shape function $\psi^{4-k}$ on $V_{4-k}$. Note that the spaces of velocity and pressure variables are not physically the same, although the same notation is used here for the sake of simplicity. 

Eqs. (19)--(23) summarize the explanations in the preceding paragraphs and depict the IFE-CGP scheme for executing one time interval of the simulation. \\
1.	Calculate {\~U}$_f^{n+1}$ on $V$ by solving
\begin{equation}
\big(\textbf{M}_v+\delta t \textbf{N} + \delta t \textbf{L}_v\big)\textrm{\~U}_f^{n+1}=\delta t\textbf{M}_v\textrm{F}^{n+1}+\textbf{M}_v\textrm{U}_f^{n}.
\end{equation}
2.	Map {\~U}$_f^{n+1}$ onto $V_{4-k}$ and obtain {\~U}$_c^{n+1}$ via
\begin{equation}
\textrm{\~U}_c^{n+1}=\textbf{R}_\textbf{4}^{\textbf{4-k}}\textrm{\~U}_f^{n+1}.
\end{equation}
3.	Calculate P$_c^{n+1}$ on $V_{4-k}$ by solving
\begin{equation}
\bar{\textbf{L}}_p\textrm{P}_c^{n+1}=\frac{\rho}{\delta t}\bar{\textbf{D}}\textrm{\~U}_c^{n+1}.
\end{equation}
4.	Remap P$_c^{n+1}$ onto $V$ and obtain P$_f^{n+1}$ via
\begin{equation}
\textrm{P}_f^{n+1}=\textbf{P}_\textbf{k}^{\textbf{k+1}}\textrm{P}_c^{n+1}.
\end{equation}
5.	Calculate U$_f^{n+1}$ via
\begin{equation}
\textbf{M}_v\textrm{U}_f^{n+1}=\textbf{M}_v\textrm{\~U}_f^{n+1}-\delta
t\textbf{G}\textrm{P}_f^{n+1}.
\end{equation}

\subsection{Computational implementation}
\label{sec:4}
A \verb C++  object oriented code is developed according to the concepts addressed in Ref. [37]. To accelerate linear algebra executions and minimize memory requirements, the standard compressed sparse row (CSR) format [38] is employed for sparse matrix-vector multiplication (SpMV) except in data transfer applications. The $ILU (0)$ preconditioned GMRES($m$) algorithm [39, 40] is chosen as an iterative linear solver. Absolute and relative tolerances are set to $10^{-8}$. The public unstructured finite element mesh generator Gmsh [41] is utilized. Calculations are performed on a single Intel(R) Xeon(R) processor with a 2.66 GHz clock rate and 64 Gigabytes of RAM.

\subsection{Computational cost analysis of IFE-CGP}
\label{sec:5}

A finite element analysis traditionally consists of three major portions: preprocessing, processing, and postprocessing. In the simulations undertaken here, the postprocessing stage mostly involves writing output files without a significant effect on the CPU time. As a result, the numerical cost, $C_c$, of the pressure correction approach on a given coarse grid is estimated by
\begin{equation}
C_c=C_{pre}+C_v+C_p,
\end{equation}
where $C_{pre}$ is the preprocessing cost, whereas $C_p$ and $C_v$ comprise, respectively, the cost of the Poisson equation (see, Eq. (17)) and the remaining algorithms in the processing portion. For transient problems with a significant number of time steps, the processing cost always overcomes the preprocessing price. By $l$-level uniform refinements of a two dimensional domain, a high-resolution simulation takes time $C_f$, roughly expressed as
\begin{equation}
C_f\approx aC_{pre}+4^lC_v+4^lC_p,
\end{equation}
with two factors $a$ and $4^l$. The factor $a$ depends on the computational resources and global matrix assembling algorithms. Because by quadrupling the $\textbf{P}_1$ element numbers, the global node numbers are doubled at minimum, $2^l$ is lower bound for the increment in the node numbers. Consequently, $4^l$ represents a lower bound associated with the matrix size enhancement. Additionally, note that $4^l$ is the lowest possible factor for cost scaling of Eqs. (16)--(17), because the cost of a matrix inversion is not linearly proportion to its size. Taking the advantages of the IFE-CGP method into account, the Poisson solution is performed on the coarse grid and its cost is not scaled. Therefore, $C_f$ is reduced to the IFE-CGP cost $C_{cgp}$, given by
\begin{equation}
C_{cgp}\approx bC_{pre}+4^lC_v+C_p+C_m,
\end{equation}
where $b$ is a new factor for the preprocessing and $C_m$ indicates the mapping expenses. The coefficient $b$ is not necessarily equal to the factor $a$ and the relation is variable. For instance, because the assembling process of $\bar{\textbf{D}}$ rather than $\textbf{D}$ is more cost-effective, it might be concluded that $a>b$. But if one takes the transpose of $\textbf{G}$ to establish $\textbf{D}$, that conclusion is questionable. Besides, the preparation of $\textbf{P}_{\textbf{k}}^{\textbf{k+1}}$ and $\textbf{R}_{\textbf{4}}^{\textbf{4-k}}$ involves an extra cost for IFE-CGP. A numerical comparison in Sect. 3.2 clarifies this point furthur. $C_m$ is shown to be negligible in comparison with the other three terms of Eq. (26) in \textbf{Remark 2.2}.

From a mesh refinement application point of view, the cost increment factor of the computational IFE-CGP tool ($h_{cgp}$) is approximated by
\begin{equation}
h_{cgp}\approx \frac{C_{cgp}}{C_c},
\end{equation}
similarly, this factor for a regular triangulation refinement ($h_f$) is conjectured to be
\begin{equation}
h_f\approx \frac{C_f}{C_c}.
\end{equation}
Based on the above discussion, $h_f$ is greater than $h_{cgp}$. This is mainly due to the factor of $4^l$ that multiples $C_p$ in Eq. (25). These results imply that mesh refinement using the CGP idea is more cost effective than the standard technique.

\textbf{Remark 2.2} \textit{(On implementing the mapping operators)}. The implementation of $\textbf{R}_{\textbf{4}}^{\textbf{4-k}}$ in the CSR format is not possible because the matrix contains null vectors. Additionally, if $\textbf{P}_{\textbf{k}}^{\textbf{k+1}}$ is constructed in the CSR format, the data prolongation cost of each IFE-CGP loop is approximated by $O(CN_f)$ where $C$ is proportional to the number of non zero elements per row of $\textbf{P}_{\textbf{k}}^{\textbf{k+1}}$ (Depending on the mesh nonuniformity, it varies between 2 and 10 in the current grids.) and $N_f$ denotes the number of pressure unknowns on  $V_{k+1}$. Though CSR is inexpensive, an easier-to-implement method is introduced next. Consider two data structures $s^p$ and $s^u$ including three ($\alpha, \beta, \gamma$) and two ($\eta, \xi$) integer indices, respectively. An array of each data structure $\{ s_{N_f}^p \}$ and $\{ s_{N_c}^u \}$, with $N_c$ equal to the node numbers on $V_k$, is created as follows:
\\
For $\forall \alpha_i \in s_i^p$ find a pair of indices $(\beta_i, \gamma_i) \in s_i^p$ that satisfies
\begin{equation}
\textrm{P}_f^{n+1}(\alpha_i)=\frac{\textrm{P}_c^{n+1}(\beta_i)+\textrm{P}_c^{n+1}(\gamma_i)}{2},
\end{equation}
and for $\forall \eta_i \in s_i^u$  find an index $\xi_i \in s_i^u$ that satisfies
\begin{equation}
 \textrm{\~{U}}_c^{n+1}(\eta_i)=\textrm{\~{U}}_f^{n+1}(\xi_i),
\end{equation}
where P$_f^{n+1}(\alpha_i)$ is the pressure value at the $\alpha_i$th node of the spanned space $V_{k+1}$, while \~{U}$_c^{n+1}(\eta_i)$ is the restricted intermediate velocity of the $\eta_i$th node of the discretized space $V_k$. P$_c^{n+1}(\beta_i)$, P$_c^{n+1}(\gamma_i)$, and \~{U}$_f^{n+1}(\xi_i)$ are similarly defined. With respect to the prolongation function, this trick improves the performance by reducing the computational effort to $O(N_f)$, with only moderately increased memory usage. Likewise, the numerical expense of the suggested injection operator is of order $O(N_c)$.



\section{Results and discussion}
\label{sec:1}
To evaluate the various aspects of the IFE-CGP method, three standard test cases are studied: the Taylor-Green decaying vortex problem, the flow over a backward-facing step, and the flow around a circular cylinder. Here, the grid resolution of a numerical simulation is denoted by $M:N$, where $M$ indicates the element numbers of a fine grid used for the advection-diffusion solver, and $N$ demonstrates the element numbers of a corresponding coarsened grid for the Poisson solver. When $N$ is equal to $M$, the standard, non-IFE-CGP, algorithm is recovered. $l$ indicates the level of mesh coarsening used in the IFE-CGP method, and is equal to zero for the standard algorithm. If function $A(\textbf{\textit{x}},t)$ is assumed to be a finite element approximation of an exact solution, $a(\textbf{\textit{x}},t)$, on the domain $V$, with $M$ elements, the difference between these two functions, $e(\textbf{\textit{x}},t)$, is defined as:
\begin{equation}
e(\textbf{\textit{x}},t)=A(\textbf{\textit{x}},t)-a(\textbf{\textit{x}},t).
\end{equation}
 The discrete norms are defined as:
\begin{equation}
\|e\|_{L^2(V)}:=\sqrt{\frac{1}{M}\sum_{i=1}^M \|e\|_{L^2(E_i)}^2},
\end{equation}
\begin{equation}
\|e\|_{L^\infty(V)}:=\max_{1\leq i \leq M}\big(\|e\|_{L^\infty(E_i)}\big) ,
\end{equation}
where $E_i$ is the discrete space of $i$th element, $E$, of $V$. Note that when an exact solution is not available, the norms are measured with reference to the standard algorithm ($l=0$).

\subsection{Taylor-Green vortex in a non trivial geometry}
\label{sec:2}
The Taylor-Green vortex problem [42] is a widely-used benchmark problem which is an exact analytic solution of the unsteady incompressible Navier-Stokes equations in two-dimensions (see e.g., [6]). The velocity field is given by:
\begin{equation}
u(x,y,t)=-\cos(2\pi x)\sin(2\pi y)\exp(-8\pi^2 \mu t),
\end{equation}
\begin{equation}
v(x,y,t)=\sin(2\pi x)\cos(2\pi y)\exp(-8\pi^2 \mu t).
\end{equation}
And the pressure field is given by:
\begin{equation}
p(x,y,t)=-\frac{\cos(4\pi x)+\cos(4\pi y)}{4}\exp(-16\pi^2 \mu t).
\end{equation}
A density value of $\rho=1$ kg/m$^3$ and a viscosity of $\mu=0.01$ Pa$\cdot$s are used.
One goal of the Taylor-Green test case is an examination of the IFE-CGP method capability in complex geometries. For this purpose, a square domain with a circular hole is chosen such that
\\
$V:=\{(x, y)| (0.25)^2 \leq (x-0.5)^2+(y-0.5)^2, 0 \leq x, y \leq 1\}.$
\\
The geometry is depicted in Fig. 2 and details of the meshes are given. A similar computational domain (a rectangular hole instead of a circular one) has been implemented by J. M. Alam et al. [43] to perform this test case. A second goal of this section is an investigation of the effects of velocity Dirichlet boundary condition (and consequently artificial Neumann boundary conditions) on the rate of accelerating computations by the IFE-CGP algorithm. Therefore, the velocity domain boundaries are set to the exact solution of Eqs. (34)--(35). These types of boundary conditions have been previously applied to the Taylor-Green vortex problem in the literature [6, 27]. San and Staples [16] have also studied this problem to validate CGPRK3 performance, but using periodic boundary conditions. In this way, an opportunity for comparison is provided. As a last concern, the effects of the prolongation operator on the thickening of artificial boundary layers are investigated. The simulations are run with a constant CFL number. As a result, a time step of 0.01 s is selected for the 3384:3384 case, and this is halved for each quadrupling of the advection-diffusion solver grid.

\begin{figure*}
\centering
\includegraphics[width=185 mm]{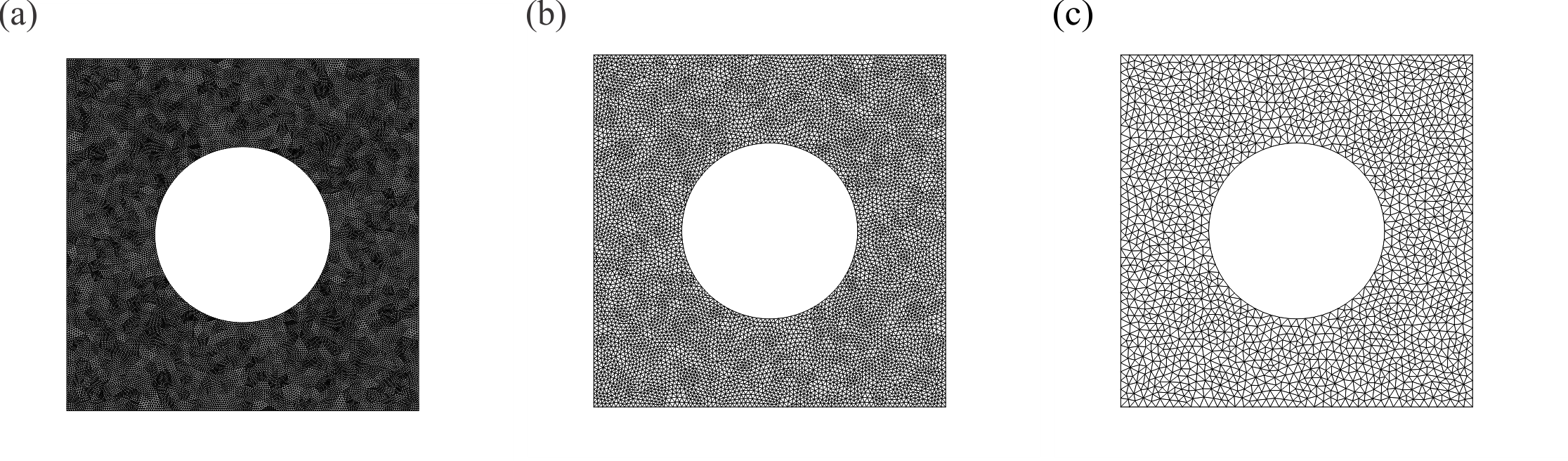}
\caption{Representation of the triangular finite element meshes used for solving Poisson’s equation in the simulation of Taylor-Green vortex. \textbf{a} After one level coarsening $(l=1)$, 27520 nodes and 54144 elements; \textbf{b} After two levels coarsening $(l=2)$, 6992 nodes and 13536 elements; \textbf{c} After three levels coarsening $(l=3)$, 1804 nodes and 3384 elements.}
\label{fig:2} 
\end{figure*}

The discrete norms of the velocity field for different mesh resolutions are tabulated in Table 1. Considering all the cases, for one and two levels ($l=1$, and $2$) of the Poisson grid coarsening, the minimum and maximum of the error percentage relative to the finest mesh ($l=0$) are, respectively, 0.30\% and 3.61\%. However, a considerable reduction in the velocity accuracy is found for three levels of coarsening. For instance, the infinity norm computed for the velocity field obtained on the 216576:3384 grid resolution indicates a 99\% reduction in the accuracy level, but it is still two orders of magnitude more accurate in comparison with the resulting data captured from the full coarse scale simulation performed on the 3384:3384 grid resolution. The speedup factors achieved range from 1.601 to 2.337. The velocity and the pressure field magnitudes for the result with three levels of coarsening are depicted in Fig. 3. The flow fields have reasonable levels of accuracy, however, dampened flows can be observed near the boundaries.

\begin{table*}
\centering
\caption{Error norms, CPU times per time step, and relative speedups for different grid resolutions of the Taylor-Green vortex simulation at $t=1$ s. Resolution in the form of $M:N$ represents the grid resolution of the advection-diffusion solver, $M$ elements, and Poisson’s equation, $N$ elements.}
\label{tab:1}   
\begin{tabular}{llllll}
\hline\noalign{\smallskip}
$l$ & Resolution & $\| \textbf{\textit{u}}\|_{L^\infty(V)}$ & $\| \textbf{\textit{u}}\|_{L^2(V)}$ & CPU time/$\delta$t & Speedup \\
\noalign{\smallskip}\hline\noalign{\smallskip}
0 & 216576:216576 & 1.59938E$-$9 & 8.34306E$-$10 & 91644000 & 1.000 \\
1 & 216576:54144 & 1.59453E$-$9 & 8.38567E$-$10 & 57223280 & 1.601 \\
2 & 216576:13536 & 1.63584E$-$9 & 8.64473E$-$10 & 40771280 & 2.247 \\
3 & 216576:3384 & 3.18962E$-$9 & 1.15333E$-$9 & 40339600 & 2.272 \\
\noalign{\smallskip}\hline\noalign{\smallskip}
0 & 54144:54144 & 1.29418E$-$8 & 6.78120E$-$9 & 1546544 & 1.000 \\
1 & 54144:13536 & 1.29274E$-$8 & 6.85697E$-$9 & 761040 & 2.032 \\
2 & 54144:3384 & 1.38146E$-$8 & 7.48644E$-$9 & 661788 & 2.337 \\
\noalign{\smallskip}\hline\noalign{\smallskip}
0 & 13536:13536 & 1.05619E$-$7 & 5.58118E$-$8 & 30944 & 1.000 \\
1 & 13536:3384 & 1.08807E$-$7 & 5.73166E$-$8 & 13598 & 2.275 \\
\noalign{\smallskip}\hline\noalign{\smallskip}
0 & 3384:3384 & 8.73658E$-$7 & 4.69697E$-$7 & 661 & 1.000 \\
\noalign{\smallskip}\hline
\end{tabular}
\end{table*}

\begin{figure*}
\centering
\includegraphics[width=179 mm]{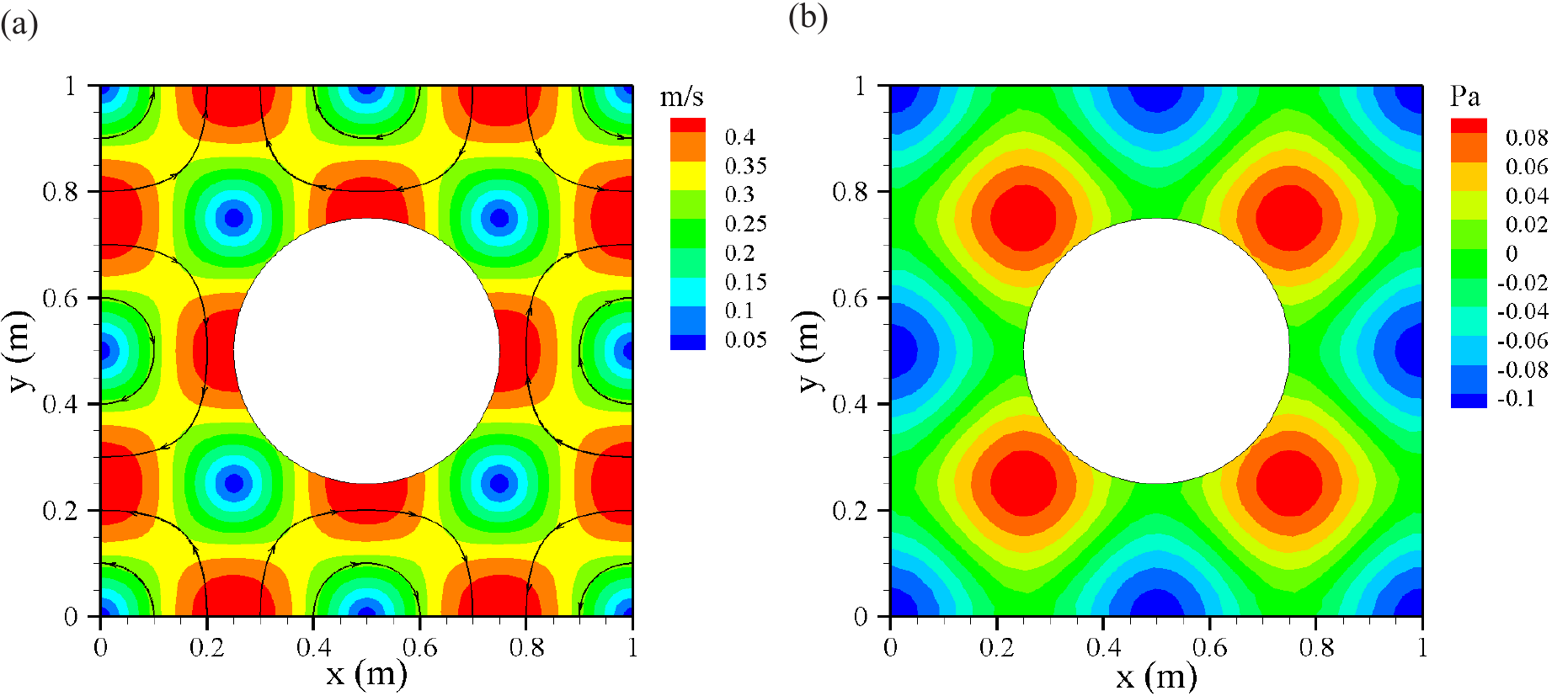}
\caption{The solution of Taylor-Green vortex in a non trivial geometry at $t=1$ s obtained using the CGP $(l=3)$ method. The resolution of the nonlinear and linear equations is, respectively, 54144 and 3384 elements. \textbf{a} Velocity magnitude contour and streamlines; \textbf{b} Pressure field.}
\label{fig:3}
\end{figure*}

In the non-incremental pressure correction methods, for a non-smooth domain with artificial Neumann boundary conditions, the pressure error norms are not sensitive to the spatial resolution and they are mostly not improved by increasing the node numbers [27]. Contrarily, the pressure gradient error norms decrease by an increase in the number of degree of freedom [27]. The data collected in Table 2 demonstrates these findings. Regarding the standard computations ($l=0$), the discrete norms of the pressure are only reduced by one order of magnitude with three levels of mesh refinement of both the pressure and the velocity spaces, whereas these quantities for the pressure gradient have a reduction of three orders of magnitude with the same grid resolution increase. By looking at the resulting pressure error norms using IFE-CGP, we find almost the same data obtained using the standard algorithm. But this does not mean that the IFE-CGP procedure conserves the accuracy of the pressure field, because the pressure field from the full fine scale simulations itself has a high level of error. In fact, it can be only concluded that the prolongation operator does not make the results worse. In contrast with the pressure norms, we observe similar trends between the velocity and pressure gradient norms for the IFE-CGP results. For instance, for one level ($l=1$) of coarsening, the minimum and maximum of the error percentage with reference to the finest mesh ($l=0$) are, respectively, 5.37\% and 16.83\%. As a note, the discrete norms calculated for the pressure gradient are relatively small in comparison with the velocity and pressure norms, and it is because we use the discrete gradient operator \textbf{G} in order to compute the gradient of both the exact solution and the numerical result. Recall that at the velocity correction step (see Eq. (15)), the pressure gradient is used to correct the velocity field, not the pressure itself. Hence, since IFE-CGP conserves the pressure gradient accuracy, the fidelity of the velocity field is also conserved.

\begin{table*}
\centering
\caption{The infinity and second error norms of pressure quantity and its gradient for the Taylor-Green vortex simulation at $t=1$ s. Resolution in the form of $M:N$ represents the grid resolution of the advection-diffusion solver, $M$ elements, and Poisson’s equation, $N$ elements.}
\label{tab:2} 
\begin{tabular}{llllll}
\hline\noalign{\smallskip}
$l$ & Resolution & $\|{p}\|_{L^\infty(V)}$ & $\|p\|_{L^2(V)}$ & $\| \textbf{G}$P$\|_{L^\infty(V)}$ & $\| \textbf{G}$P$\|_{L^2(V)}$ \\
\noalign{\smallskip}\hline\noalign{\smallskip}
0 & 216576:216576 & 1.34241E$-$6 & 6.00500E$-$7 & 3.21871E$-$12 & 1.81154E$-$13 \\
1 & 216576:54144 & 1.34241E$-$6 & 6.00416E$-$7 & 3.09665E$-$12 & 1.90897E$-$13 \\
2 & 216576:13536 & 1.34241E$-$6 & 6.00270E$-$7 & 2.22832E$-$12 & 2.43972E$-$13 \\
3 & 216576:3384 & 1.34241E$-$6 & 6.00171E$-$7 & 3.65721E$-$12 & 5.78906E$-$13 \\
\noalign{\smallskip}\hline\noalign{\smallskip}
0 & 54144:54144 & 4.98048E$-$6 & 2.38748E$-$6 & 4.09860E$-$11 & 3.48936E$-$12 \\
1 & 54144:13536 & 4.98048E$-$6 & 2.38692E$-$6 & 3.96108E$-$11 & 3.81608E$-$12 \\
2 & 54144:3384 & 4.98048E$-$6 & 2.38657E$-$6 & 3.22830E$-$11 & 6.43534E$-$12 \\
\noalign{\smallskip}\hline\noalign{\smallskip}
0 & 13536:13536 & 1.89334E$-$5 & 9.42727E$-$6 & 4.86969E$-$10 & 7.21105E$-$11 \\
1 & 13536:3384 & 1.89334E$-$5 & 9.42590E$-$6 & 5.31351E$-$10 & 8.42464E$-$11 \\
\noalign{\smallskip}\hline\noalign{\smallskip}
0 & 3384:3384 & 6.74979E$-$5 & 3.66242E$-$5 & 6.55823E$-$9 & 1.66480E$-$9 \\
\noalign{\smallskip}\hline
\end{tabular}
\end{table*}

The Cartesian coordinate of an element’s center at which the infinity error norm occurs for the velocity, pressure, and pressure gradient in the Taylor-Green vortex domain are tabulated in Table 3. For different levels of coarsening, the location changes (except for the pressure error, which remains unchanged). However, it is always near the boundaries of either the ring or the square. In particular, the maximum errors for the velocity and pressure gradient fields occur on the edge of the square for ($l=3$). Thus, even using the prolongation operator of the IFE-CGP method, the maximum errors still occur around the boundaries with velocity Dirichlet conditions.

\begin{table*}
\centering
\caption{Comparison of various locations of velocity magnitude, pressure, and gradient pressure infinity-norm errors for the Taylor-Green vortex problem at $t=1$ s. $(x, y)$ indicates the center of an element, which the errors occur in. Resolution in the form of $M:N$ demonstrates the grid resolution of the advection-diffusion solver,  $M$ elements, and Poisson’s equation, $N$ elements.}
\label{tab:3} 
\begin{tabular}{lllll}
\hline\noalign{\smallskip}
&  & $\|\textbf{\textit{u}}\|_{L^\infty(V)}$ & $\|p\|_{L^\infty(V)}$ & $\|$\textbf{G}P$\|_{L^\infty(V)}$ \\
\noalign{\smallskip}
$l$ & Resolution & $(x, y)$ & $(x, y)$ & $(x, y)$ \\
\noalign{\smallskip}\hline\noalign{\smallskip}
0 & 216576:216576 & (0.576623, 0.808348) & (0.482527, 0.110053) & (0.340604, 0.695205) \\
1 & 216576:54144 & (0.580109, 0.809028) & (0.482527, 0.110053) & (0.659396, 0.304795) \\
2 & 216576:13536 & (0.601023, 0.813106) & (0.482527, 0.110053) & (0.306953, 0.664904) \\
3 & 216576:3384 & (0.770613, 0.977284) & (0.482527, 0.110053) & (0.004544, 0.243629) \\
\noalign{\smallskip}\hline
\end{tabular}
\end{table*}

From a general point of view, the spatial discretization of the advection-diffusion domain acts as a low-pass filter on the grid, and the Poisson solver also acts as a pre-filtering process [16]. The CGP procedure specifically uses the belief in order to increase saving in computational time without negatively affecting the properly-resolved advection-diffusion field. A visual demonstration of these effects is displayed in Figs. 4--5. If the difference between the exact solution and the numerical result is considered as a noise over the domain, Figure 4 and Figure 5 show the noise distribution, respectively, for the pressure and velocity variables. For example, by switching the grid resolution from 216576:216576 (see Fig. 4a and Fig. 5a) to 216576:54144 (see Fig. 4b and Fig. 5b), the maximum absolute value of the pressure noise roughly changes from $7 \times 10^{-3}$ to $7.5 \times 10^{-3}$, implying a 7.14\% noise increment, whereas there is no significant noise enhancement for the velocity field. Similar behavior is observed for further levels of the Poisson grid coarsening. For instance, as it can be seen in Fig. 4d and Fig. 5d, the noise of the pressure and velocity domains with three levels of coarsening increases, respectively, to 85.71\% and 33.33\%, demonstrating that the low-pass filtering feature of the discretization of the advection-diffusion part of the algorithm is why the large errors of the pressure field do not get transmitted to the velocity field in CGP methods.

\begin{figure*}
\centering
\includegraphics[width=175 mm]{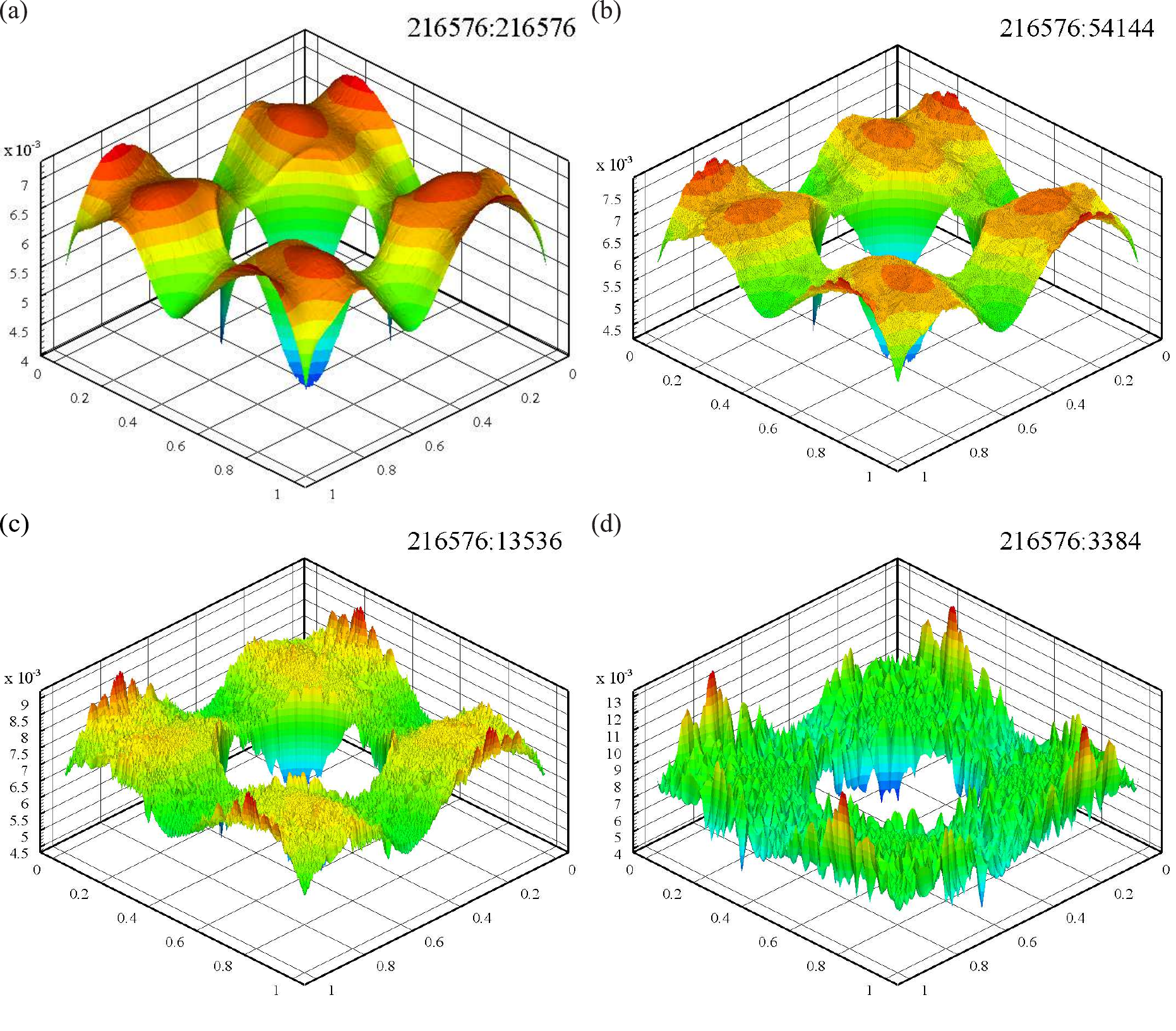}
\caption{Distribution of point wise pressure error for Taylor-Green vortex problem at $t=1$ s. Labels in the form of $M:N$ indicate the grid resolution of the advection-diffusion solver, $M$ elements, and the Poisson equation, $N$ elements. \textbf{a} Regular computation $(l=0)$; \textbf{b} IFE-CGP $(l=1)$; \textbf{c} IFE-CGP $(l=2)$; \textbf{d} IFE-CGP $(l=3)$.}
\label{fig:4} 
\end{figure*}

\begin{figure*}
\centering
\includegraphics[width=175 mm]{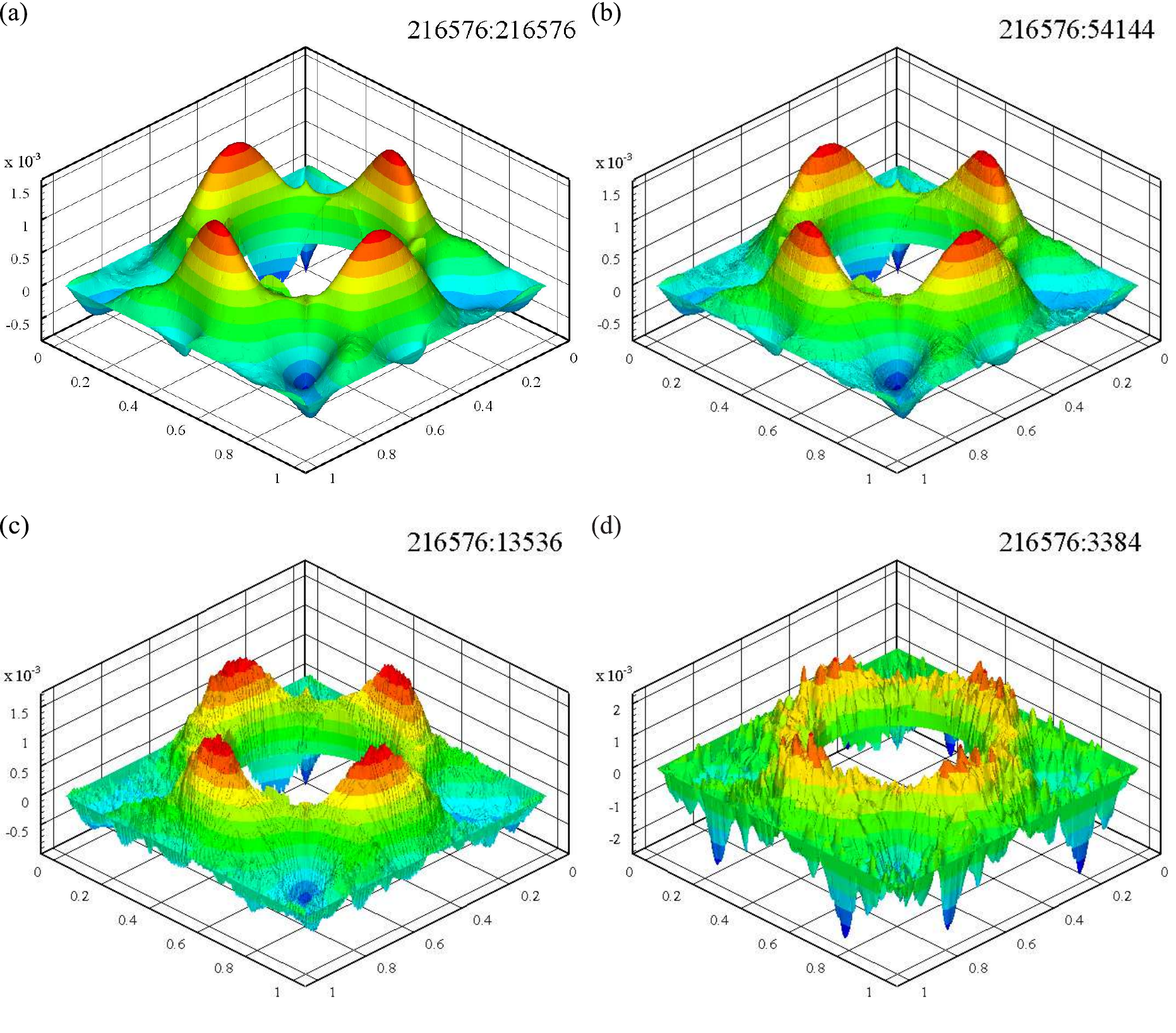}
\caption{Distribution of point wise velocity magnitude error for Taylor-Green vortex problem at $t=1$ s. Labels in the form of $M:N$ indicate the grid resolution of the advection-diffusion solver, $M$ elements, and the Poisson equation, $N$ elements. \textbf{a} Regular computation $(l=0)$; \textbf{b} IFE-CGP $(l=1)$; \textbf{c} IFE-CGP $(l=2)$; \textbf{d} IFE-CGP $(l=3)$.}
\label{fig:5} 
\end{figure*}

\textbf{Remark 3.1} \textit{(On the error order of the non incremental pressure-correction scheme)}. By calculating the slope of the discrete norms for the standard algorithm results, it is induced that the order of the overall accuracy is not fully a function of the spatial resolution. This is because the time error is dominant over the spatial step in the non-incremental pressure correction scheme with Dirichlet boundary conditions [1]. Numerically, in order to remove the time error, very small time intervals should be taken. For instance, this time error becomes invisible if the time step is chosen to be $\delta t=10^{-7}$ s for the simulation with a 3384:3384 spatial resolution. This severe condition is not odd because the fluid domain is surrounded by velocity Dirichlet boundaries from both the outside and inside. However, our computational resources do not allow us to choose such a tiny time increment for this test problem. Note that this discussion is not directly relevant to the topic of CGP methodology, but it can be concluded that a multigrid method (e.g., the CGP technique) provides optimal results in the presence of velocity Dirichlet boundary conditions if users take the incremental pressure-correction schemes introduced in Refs [1, 6, 27, 44, and 45].

\subsection{Flow over a backward-facing step}
\label{sec:3}
To study the IFE-CGP algorithm efficiency in the presence of open boundary conditions, the flow over a backward facing step inside a channel is analyzed. Fig. 6 presents the problem geometry and imposed boundary conditions. Because an inlet channel upstream significantly affects the flow simulation at low Reynolds number [46], the inflow boundary is located at the step and is described by a parabolic profile:
\begin{equation}
u=24.0(y-0.5)(1.0-y),
\end{equation}
\begin{equation}
v=0.
\end{equation}
The origin of the coordinate system is placed in the lower left corner of the step. Homogeneous natural Neumann conditions
\begin{equation}
\mu \nabla \textbf{\textit{u}} \cdot \textbf{\textit{n}}=0,
\end{equation}
are enforced at the exit. The Reynolds number is calculated as
\begin{equation}
Re = \frac{\rho H\bar{u}}{\mu},
\end{equation}
where $H$ is the channel height and $\bar{u}$ represents the space-averaged mean entrance flow velocity.

\begin{figure*}
\centering
\includegraphics[width=155 mm]{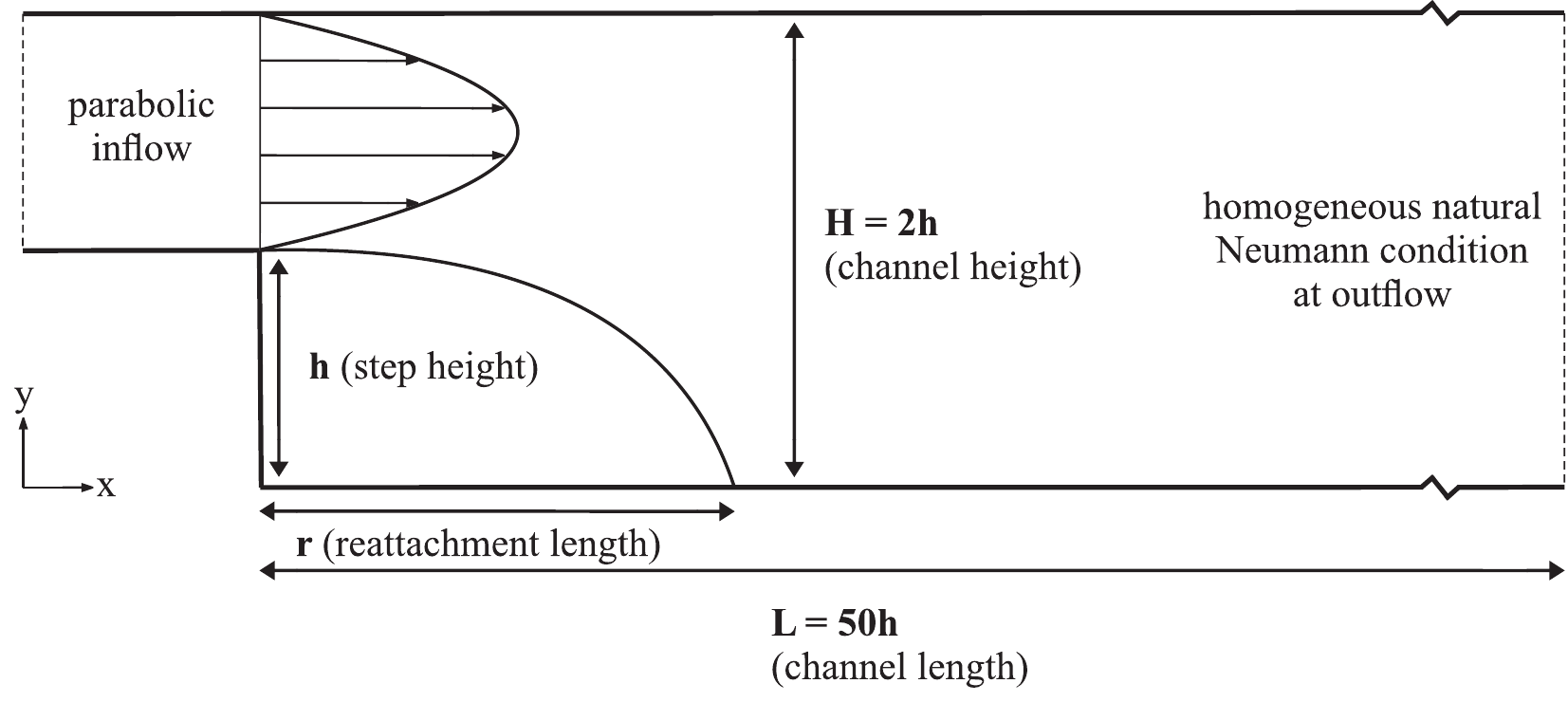}
\caption{Schematic view of flow past a backward-facing step.}
\label{fig:6}
\end{figure*}

Although the stress free boundary condition is less restrictive than Dirichlet type boundary conditions at the outflow [27], it leads to the increased the possibility of a loss in spatial accuracy [6]. This unfavorable situation is mainly due to imposing artificial homogenous Dirichlet boundary conditions in the pressure Poisson solver [6]. On the other hand, because the IFE-CGP technique reduces the number of pressure unknowns at the outflow, it is one of the current research interests to check whether the IFE-CGP methodology provides a valid solution. For this purpose, the prediction of the reattachment length $r$ with respect to $Re$ is plotted in Fig. 7. The obtained results for one ($l=1$) and two ($l=2$) levels of coarsening reveal good agreement with the numerical data of Kim and Moin [47], and Erturk [48]. At $Re=800$, the reattachment length just differs from that reported by Erturk [48] about 0.7\% and 4.0\%, respectively, for the IFE-CGP ($l=1$) and IFE-CGP ($l=2$) algorithms. However, the IFE-CGP ($l=3$) approach vastly overestimates the reattachment length after the Reynolds number has reached the value 300.

\begin{figure}
\centering
\includegraphics[width=84 mm]{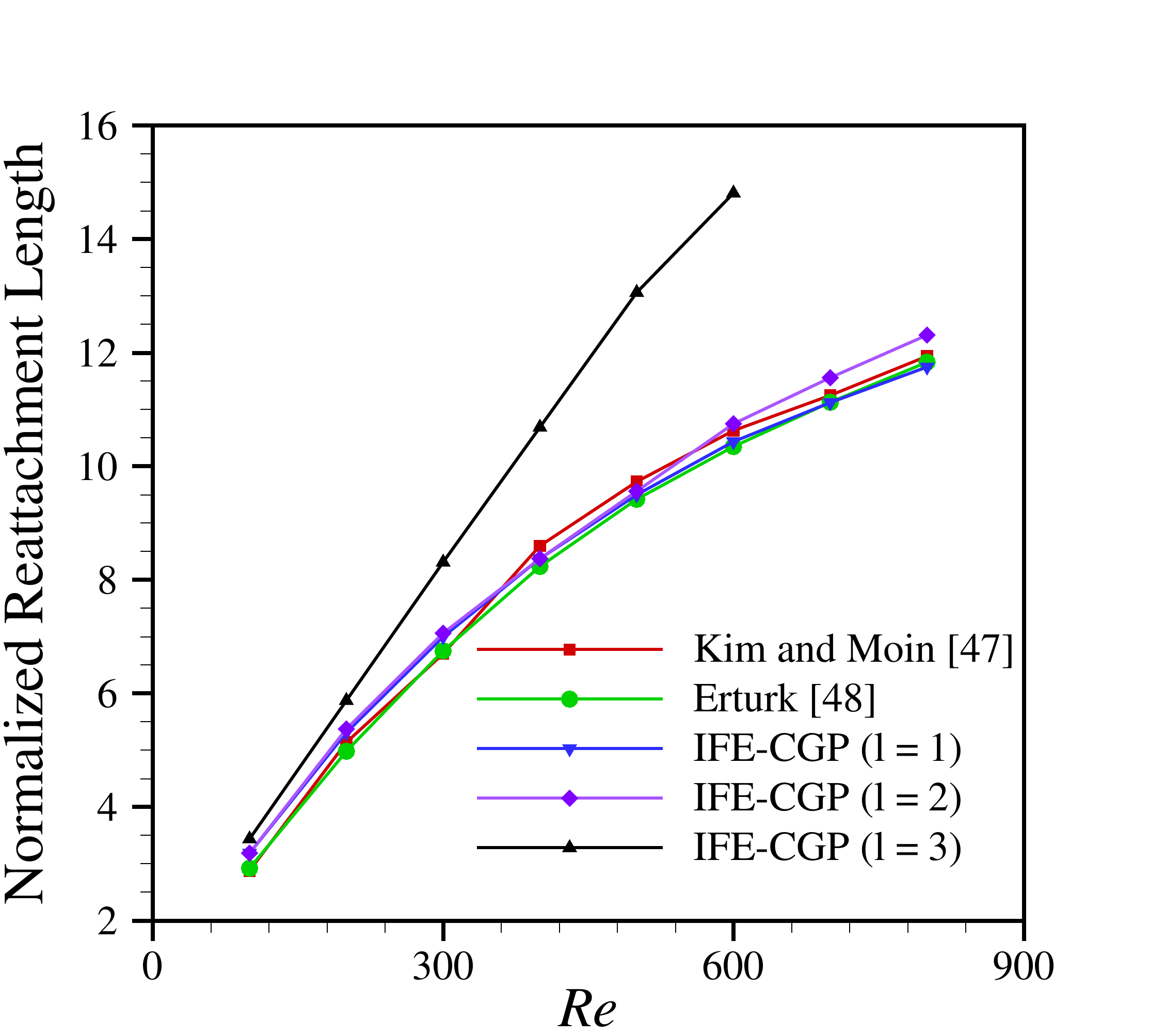}
\caption{Normalized reattachment length $(r/h)$ as a function of Reynolds number.}
\label{fig:7} 
\end{figure}

To save space, detailed results are only presented for a Reynolds number of $Re=800$. The time step is chosen to be $\delta t=0.035$ s. Based on our numerical experiments, the flow field reaches stationarity at $t=280$ s for the regular fine computations. The results for all other options are reported at this fluid flow time.

Figure 8 depicts the axial velocity contour maps of the flow simulated using both the normal and the IFE-CGP processes. Additionally, the efficiency and accuracy of the velocity field for the standard algorithm and the IFE-CGP approach are compared in Table 4. Flow velocity variables with one and two levels of coarsening agree well with the fine scale standard computations. Significantly, a 30-times reduction in computational cost of the IFE-CGP solution with the 102400:1600 resolution is obtained, as the corresponding velocity error norms are still in the acceptable range. As can be seen from Fig. 9a, interestingly, the advection-diffusion solver diverges for a simulation with the 1600:1600 resolution. This behavior will be discussed comprehensively in Sect. 3.4.

\begin{table*}
\centering
\caption{Comparison of relative norm errors and total CPU times between the standard and IFE-CGP algorithms for the flow past a backward-facing step at $Re=800$.}
\label{tab:4}
\begin{tabular}{llllll}
\hline\noalign{\smallskip}
$l$ & Resolution & $\|\textbf{\textit{u}}\|_{L^\infty(V)}$ & $\|\textbf{\textit{u}}\|_{L^2(V)}$ & CPU (s) & Speedup\\
\noalign{\smallskip}\hline\noalign{\smallskip}
0 & 102400:102400 & - & - &3447941&1.000\\
1 & 102400:25600 & 6.69447E$-$6 &7.88459E$-$7&173052&19.924 \\
2 & 102400:6400 & 1.43997E$-$5 &4.87393E$-$6&120500&28.613 \\
3 & 102400:1600 & 4.4854E$-$5 & 2.23362E$-$5&116745&29.534  \\
\noalign{\smallskip}\hline
\end{tabular}
\end{table*}

\begin{figure*}
\centering
\includegraphics[width=140 mm]{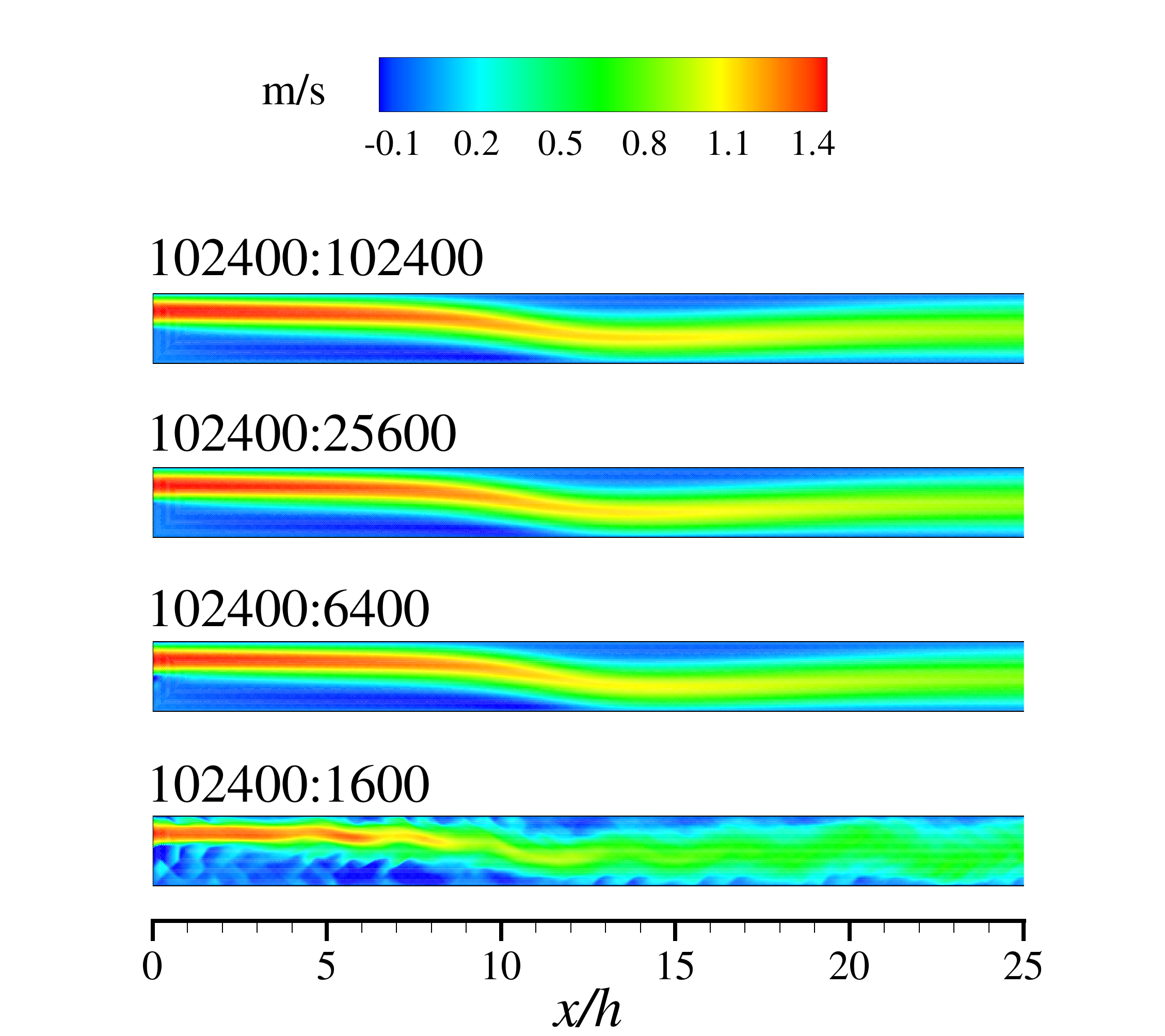}
\caption{Horizontal velocity component contour plot of flow over backward-facing step at $Re=800$. Labels in the form of $M:N$ indicate the grid resolution of the advection-diffusion solver, $M$ elements, and Poisson's equation, $N$ elements.}
\label{fig:8} 
\end{figure*}

\begin{figure*}
\centering
\includegraphics[width=140 mm]{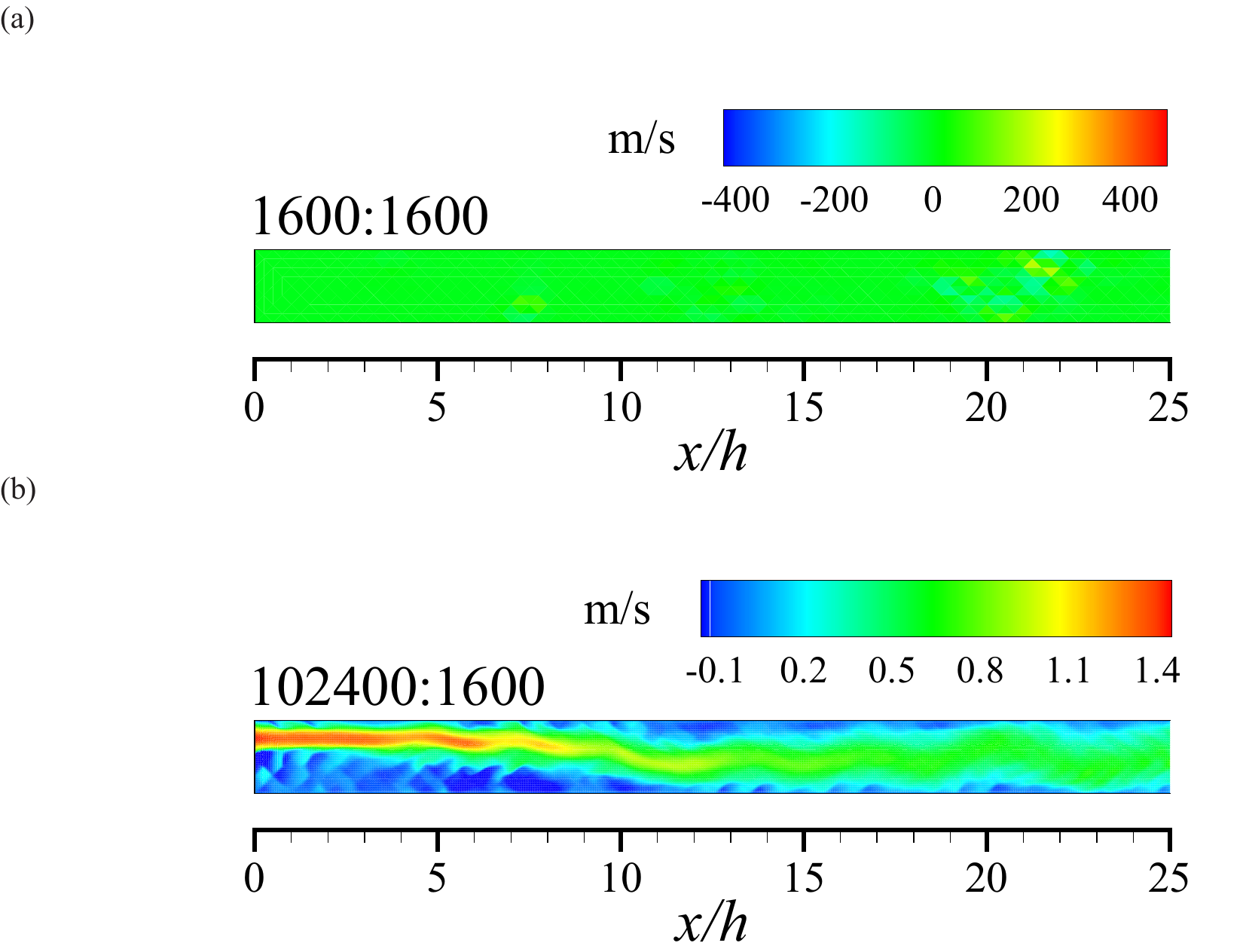}
\caption{Demonstration of partial mesh refinement application of the CGP method, a comparison between axial velocity contours of \textbf{a} Regular computation, diverged, and \textbf{b} The IFE-CGP $(l=3)$, converged, for flow past backward-facing step at $Re=800$. Labels in the form of $M:N$ indicate the grid resolution of the advection-diffusion solver, $M$ elements, and the Poisson equation, $N$ elements.}
\label{fig:9}
\end{figure*}

The CPU times devoted to the restriction/prolongation operators are tabulated in Table 5. By increasing the coarsening level $l$, the prolongation operator becomes more expensive, whereas the time consumed by means of the restriction function decreases. To explain this fact, let's consider, for instance, the data mapping procedure of the IFE-CGP $(l=3)$ strategy on the 102400:1600 grid resolution. The restriction operator directly injects the intermediate velocity field data from a grid with 102400 elements into the corresponding coarse grid with 1600 elements. From a programming point of view, this operation needs only 905 loops, which is the pressure node numbers of the coarse grid. The prolongation operator, in contrast, has to utilize two intermediate grids, associated with $l=2$ and $l=1$, in order to extrapolate the pressure domain data. To handle this transformation, the prolongation operation requires 68659 loops, which is the summation of grid points belonged to the two intermediate as well as the finest meshes. As can be seen from Table 5, the IFE-CGP method slightly increases the time spent on the preprocessing block. Even though the $\bar{\textbf{L}}_p$ and $\bar{\textbf{D}}$ matrix assembling process is computationally cheaper in comparison with the standard algorithm, the mapping operator constructions ultimately overcome these savings. That is, the ratio $b/a$ is greater than 1.00 in Table 4. Appling AMG tools in order to establish the $\bar{\textbf{L}}_p$ and $\bar{\textbf{D}}$ matrices in the IFE-CGP technique is a way to optimize the preprocessing subroutine costs.

\begin{table*}
\centering
\caption{CPU times consumed by restriction and prolongation operators, preprocessing segment, and its ratio ($b/a$) in comparison between the IFE-CGP and standard schemes for the backward-facing step flow simulation (see Sect. 2.4 for further details).}
\label{tab:5}
\begin{tabular}{llllll}
\hline\noalign{\smallskip}
$l$ & Resolution & Restriction (s) & Prolongation (s) & Preprocessing (s) & ($b/a$)\\
\noalign{\smallskip}\hline\noalign{\smallskip}
0 & 102400:102400 & - & - &407.471&1.000\\
1 & 102400:25600 & 0.89 & 3.11 &698.441&1.714\\
2 & 102400:6400 & 0.21 & 3.67&713.521&1.751\\
3 & 102400:1600 & 0.05 & 3.64&725.521&1.780\\
\noalign{\smallskip}\hline
\end{tabular}
\end{table*}

\subsection{Flow past a circular cylinder}
\label{sec:4}

The concern of this section is a study of the effect of curvature on the IFE-CGP method. Although this capability of the method has been already investigated for decaying vortices in curved geometries in Sect. 3.1, an external unsteady flow over a cylinder is a more physically meaningful benchmark case [49]. San and Staples [16] have performed this fluid mechanics problem by means of the CGPRK3 solver, but for steady-state flows, at $Re=40$, and exclusively using one level of coarsening ($l=1$). Here, the IFE-CGP methodology with three levels of coarsening is applied to this canonical flow problem at $Re=100$.

The computational field is considered as a rectangular domain $V:=[0, 38]\times[0, 32]$. A circle with diameter $d$ represents the cylinder in two dimensions, and its center lies at the point (8, 16). A free stream velocity $u_\infty$ parallel to the horizontal axis is imposed at the inflow boundary. The circle is treated as a rigid boundary and no-slip conditions are enforced. The velocity at the top and bottom of $V$ is perfectly slipped in the horizontal direction. The outflow velocity is specified with a natural Neumann condition, Eq. (39). The Reynolds number is determined as
\begin{equation}
Re=\frac{\rho du_\infty }{\mu}.
\end{equation}
To set this dimensionless number to $Re=100$, the density, cylinder diameter, and free stream velocity are set to 1.00; and the viscosity is set to 0.01 in the International Unit System. The described geometry and boundary conditions are taken from the literature [50, 51] to satisfy far-field assumptions. A fixed time increment of $\delta t = 0.05$ s is selected and the numerical experiments are executed until time $t=150$ s. The grids utilized by the Poisson solver for $l=0$, $l=1$, $l=2$, and $l=3$ are like those shown in Fig. 10, with 108352 nodes and 215680 elements, 27216 nodes and 53920 elements, 6868 nodes and 13480 elements and 1749 nodes and 3370 elements, respectively, with very fine grid spacing near the circle.

\begin{figure*}
\centering
\includegraphics[width=185 mm]{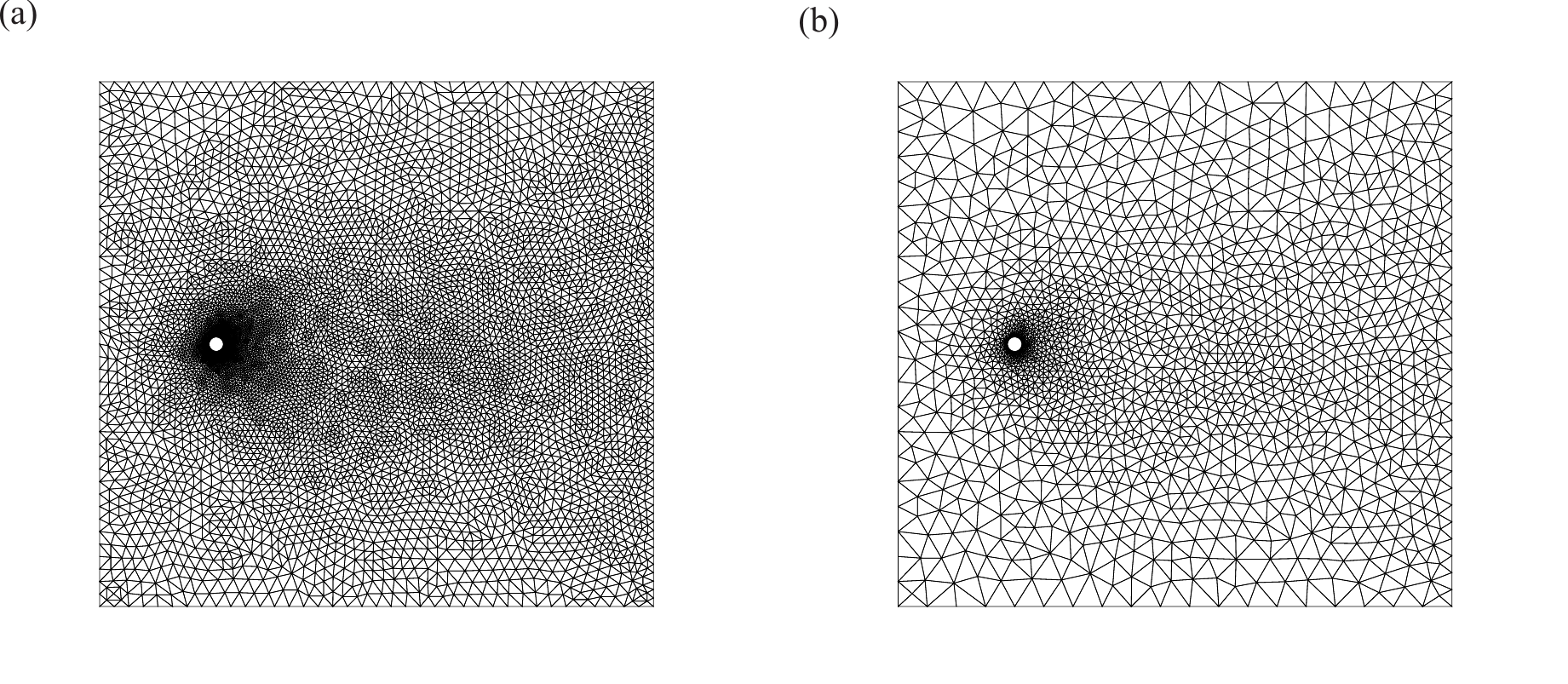}
\caption{Computational mesh for the Poisson equation solution of the flow past a circular cylinder. \textbf{a} After two levels coarsening $(l=2)$; \textbf{b} After three levels coarsening $(l=3)$. Details of the grids are reported.}
\label{fig:10}
\end{figure*}

A visual comparison between the obtained vorticity fields with and without the IFE-CGP method is made in Fig. 11 for several spatial resolutions at time $t=150$ s. For all levels of coarsening, the IFE-CGP field provides more detailed data compared to that modeled with a full coarse grid resolution. A comparison between the resulting vorticity fields with the resolutions of 215680:215680 and 13480:13480 demonstrate that the phases of periodic variation of these two fields are not equal to each other. Conversely, the fields computed by 215680:215680 and 215680:13480 oscillate with the same phase. However, there is a phase lag between the outcomes with 215680:215680 and 215680:53920 or 215680:3370 mesh resolutions. Our numerical experiments show that the phase lag between the standard and IFE-CGP results depends on the time step chosen. For instance, the velocity field phases, and consequently the vorticity ones, are the same in both the simulations performed by IFE-CGP ($l=0$) and IFE-CGP ($l=1$) tools when $\delta t=0.1$ s.

\begin{figure*}
\centering
\includegraphics[width=182 mm]{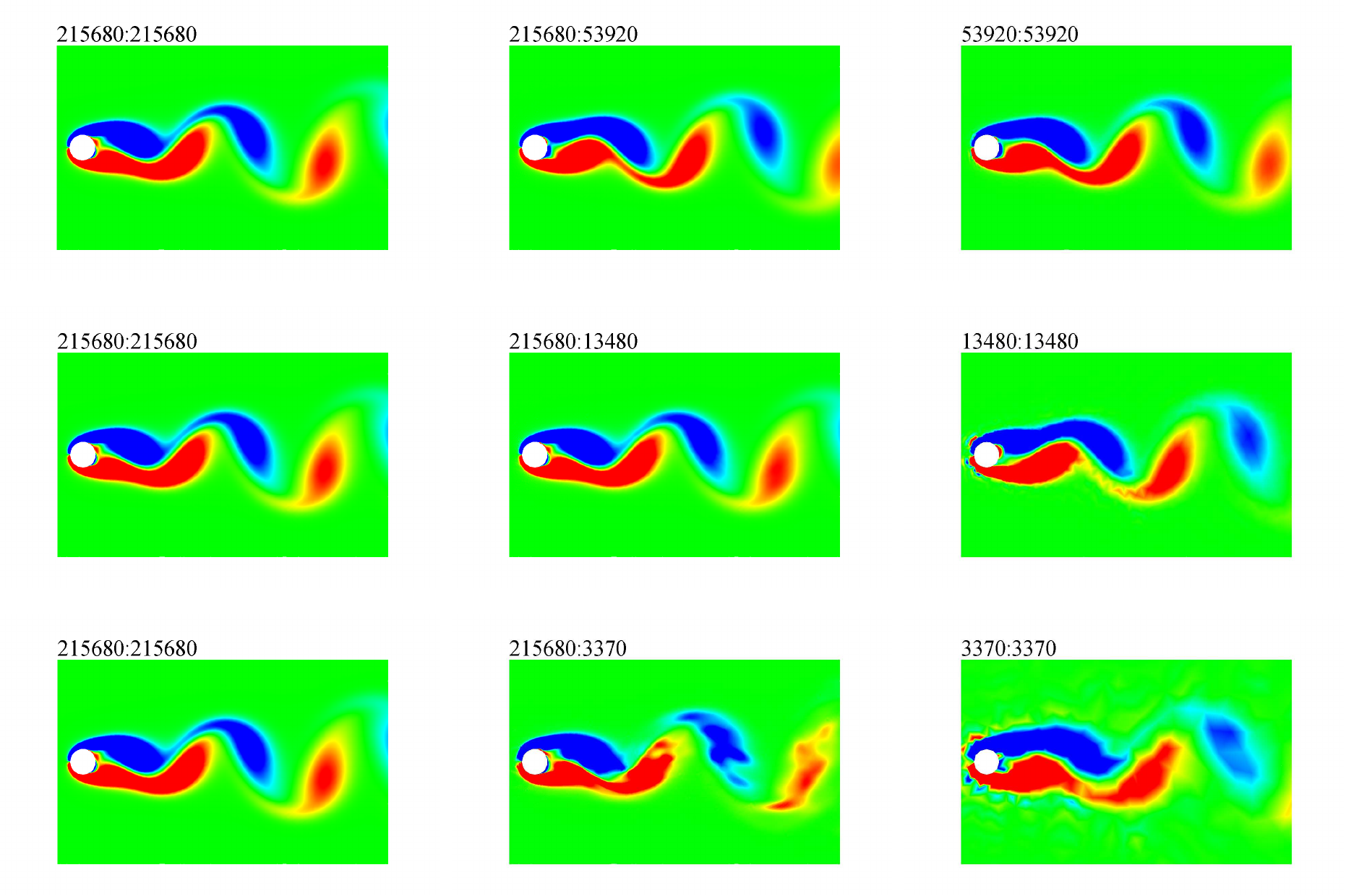}
\caption{Vorticity fields for the flow past a circular cylinder at $t=150$ s. Labels in the form of $M:N$ illustrate the spatial resolution of the advection-diffusion grid, $M$ elements, and the Poisson equation mesh, $N$ elements.}
\label{fig:11}
\end{figure*}

To more precisely analyze the IFE-CGP algorithm’s performance, velocity and vorticity distributions along the horizontal centerline, behind the cylinder and in the wake region, are shown in Fig. 12 at time $t=150$ s. By coarsening the advection-diffusion mesh at the exit of the fluid domain, the results of the 53920:53920 resolution includes spurious fluctuations. These fluctuations become stronger in the pure coarse case with the 13480:13480 resolution. However, they are successfully removed using the IFE-CGP approach. The computational times for the performed simulations are: 491068.2 s, 70809.2 s, 65492.7 s, and 56251.1 s, respectively for $l=0$  (215680:215680), $l=1$ (215680:53920), $l=2$ (215680:13480), and $l=3$  (215680:3370), leading to speedup factors between 6.935 and 8.730.

\begin{figure*}
\centering
\includegraphics[width=165 mm]{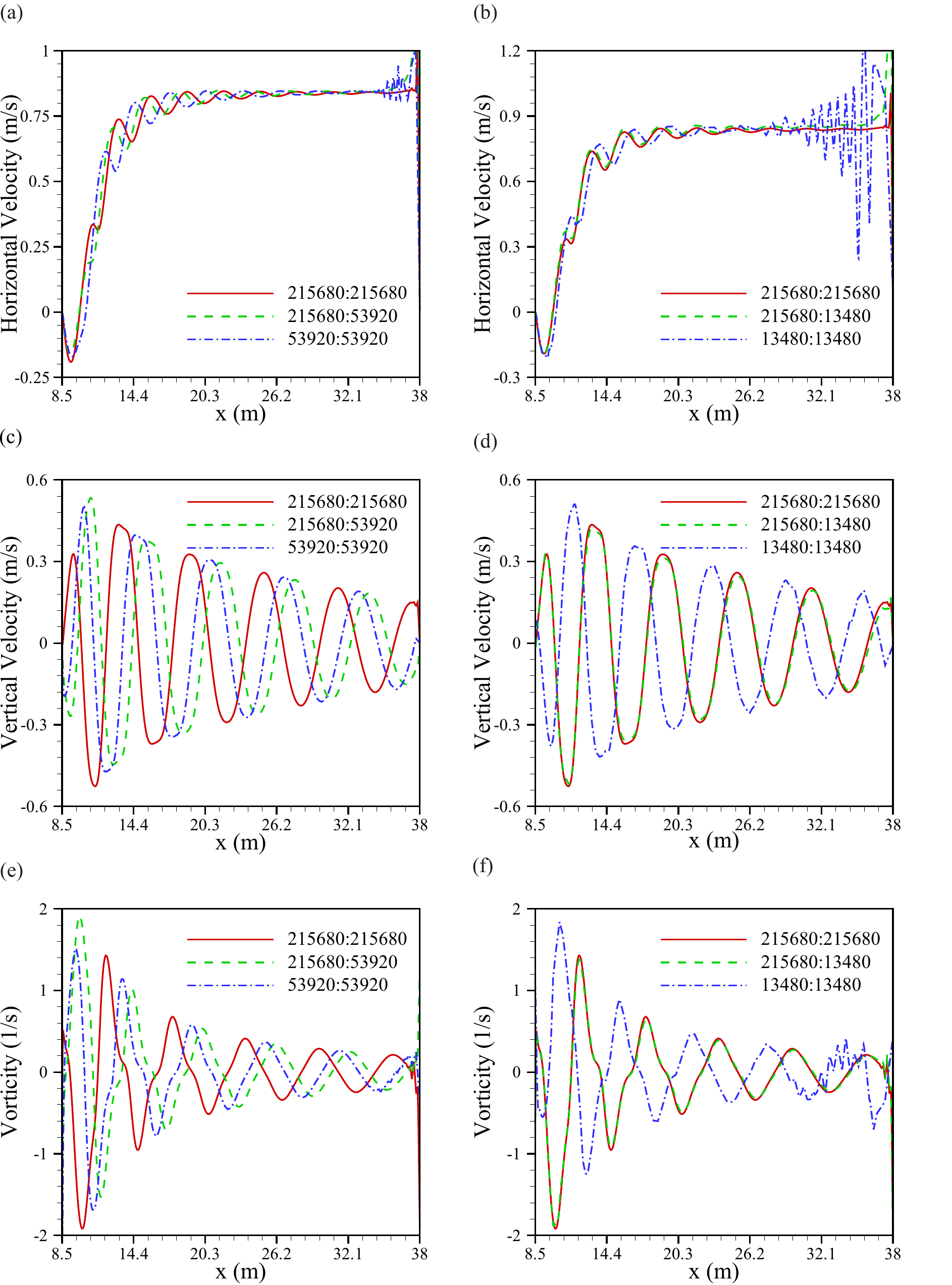}
\caption{Comparison of various output variables in the wake of the flow over a cylinder at $t=150$ s for different values of the advection-diffusion and the Poisson grid resolutions. \textbf{a} Horizontal velocity for IFE-CGP ($l=0$, 1, and 0); \textbf{b} Horizontal velocity for IFE-CGP ($l=0$, 2, and 0); \textbf{c} Vertical velocity for IFE-CGP ($l=0$, 1, and 0); \textbf{d} Vertical velocity for IFE-CGP ($l=0$, 2, and 0); \textbf{e} Vorticity for IFE-CGP ($l=0$, 1, and 0); \textbf{f} Vorticity for IFE-CGP ($l=0$, 2, and 0). The grid resolution in the form of $M:N$ shows the element numbers of the advection-diffusion grid by $M$, and the Poisson grid by $N$.}
\label{fig:12}
\end{figure*}

Table 6 lists the calculated Strouhal number ($St$), drag ($C_D$) and lift ($C_L$)  coefficients compared with the experimental and numerical studies presented in [52--55]. The Strouhal number is based on the time evolution of the blunt body lift, and is formulated as:
\begin{equation}
St=\frac{f_sd}{u_\infty},
\end{equation}
where $f_s$ is the shedding frequency. The data presented in Table 6 demonstrates that as the coarsening level ($l$) increases, the drag and lift forces slightly decrease and increase, respectively. However, they still agree well with the values found in the literature. Because the IFE-CGP method computes the pressure variable on a coarsened mesh, there is a concern it might reduce the accuracy of the drag and lift coefficients. To check this issue, the time evolution of the viscous ($C_{Lf}$), pressure ($C_{Lp}$), and total lift coefficients are plotted separately in Fig. 13. Interestingly, the viscous and pressure lift diagrams for the full fine scale and the IFE-CGP ($l=1$) simulations become nearly identical after approximately time $t=100$ s; nevertheless, the IFE-CGP configuration provides the maximum magnitude of the lift force two periods of the flow cycle earlier. Even though a numerical computation performed on a pure coarse grid degenerates the viscous lift coefficient, choosing one level of coarsening for the IFE-CGP mechanism does not influence the viscous force integrity. For two further levels of the Poisson grid coarsening, spurious fluctuations can be observed at the beginning of the fluid flow simulation. For the IFE-CGP ($l=3$) results, when the simulation is using the 215680:3370 grid resolution, the lift coefficient reaches its final amplitude at approximately time $t=40$ s, roughly 60 s earlier than the standard full fine scale computations. Surprisingly, the lift force obtained by full coarse scale (3370:3370) computations oscillates around approximately 0.2 (instead of 0.0), showing the flow field is under-resolved. Contrarily, the IFE-CGP ($l=3$) lift force is more reliable since it fluctuates around 0.0.

\begin{table}
\caption{Strouhal number, drag and lift coefficients for $Re=100$.}
\label{tab:6}       
\begin{tabular}{llll}
\hline\noalign{\smallskip}
Study & $St$ & C$_D$ & C$_L$  \\
\noalign{\smallskip}\hline\noalign{\smallskip}
Braza et al. [52] & 0.160 & 1.364$\pm$0.015 & $\pm$0.25\\
Liu et al. [53] & 0.165 & 1.350$\pm$0.012 & $\pm$0.339 \\
Hammache and Gharib [54] & 0.158 & - & - \\
Rajani et al. [55] & 0.156 & - & - \\
IFE-CGP ($l=1$) & 0.156 & 1.258$\pm$0.006 & $\pm$0.217 \\
IFE-CGP ($l=2$) & 0.152 & 1.223$\pm$0.006 & $\pm$0.250 \\
IFE-CGP ($l=3$) & 0.149 & 1.168$\pm$0.034 & $\pm$0.344 \\
\noalign{\smallskip}\hline
\end{tabular}
\end{table}

\begin{figure*}
\centering
\includegraphics[width=165 mm]{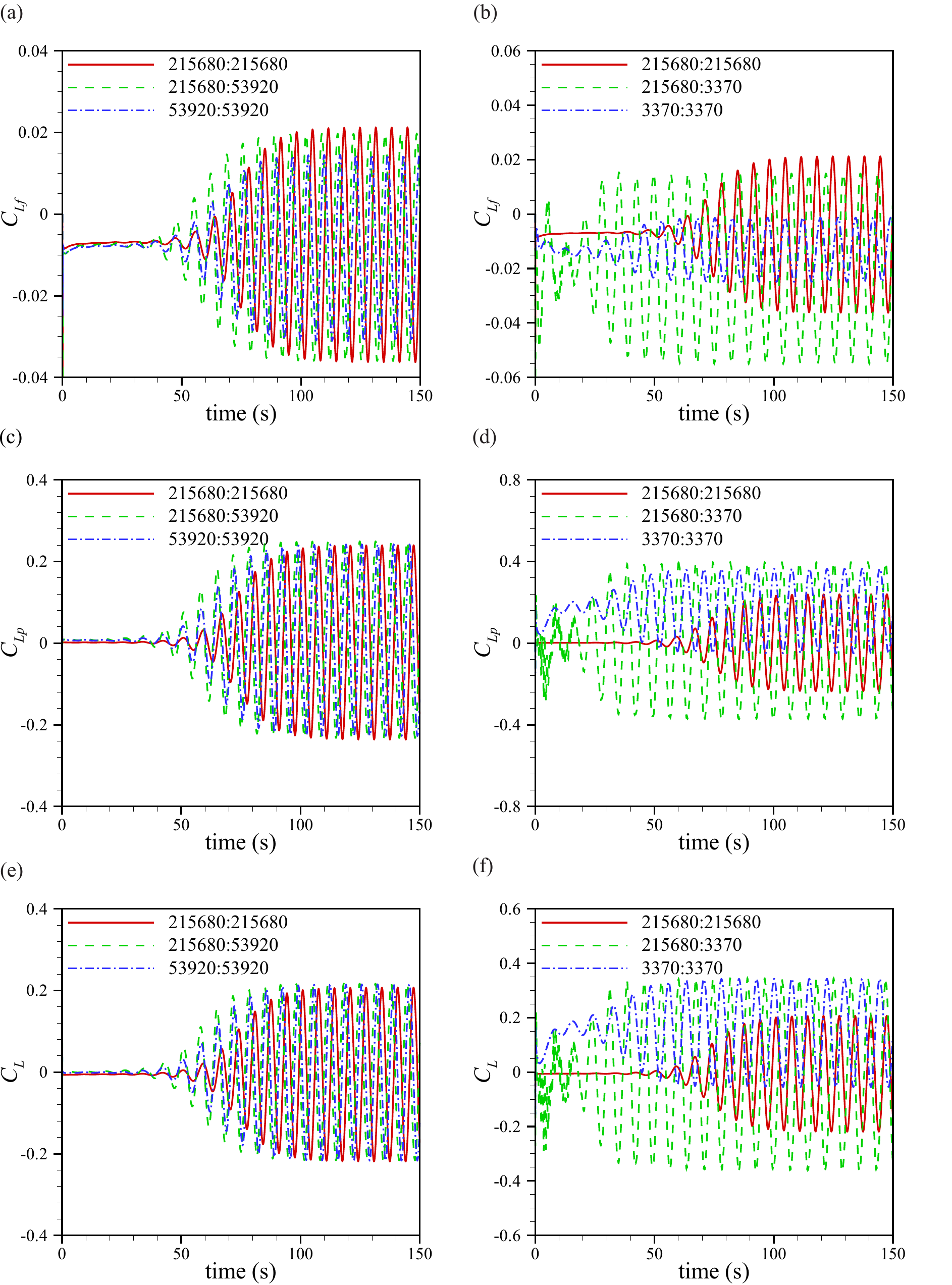}
\caption{A Comparison between the time evolution of lift coefficients of the flow past a cylinder at $Re=100$ for different combinations of the advection-diffusion and the Poisson mesh resolutions. \textbf{a} Viscous lift coefficient for IFE-CGP ($l=0$, 1, and 0); \textbf{b} Viscous lift coefficient for IFE-CGP ($l=0$, 3, and 0); \textbf{c} Pressure lift coefficient for IFE-CGP ($l=0$, 1, and 0); \textbf{d} Pressure lift coefficient for IFE-CGP ($l=0$, 3, and 0); \textbf{e} Lift coefficient for IFE-CGP ($l=0$, 1, and 0); \textbf{f} Lift coefficient for IFE-CGP ($l=0$, 3, and 0). The spatial resolution in the format of $M:N$ indicates the finite element numbers of the advection-diffusion grid by $M$, and the Poisson grid by $N$.}
\label{fig:13}
\end{figure*}

The magnitude of the centerline pressure and its gradient along the $x$-axis in the wake region are shown in Fig. 14 for the full fine (215680:215680), IFE-CGP ($l=1$, and 2) (215680:53920 and 215680:13480), and full coarse (53920:53920 and 13480:13480) simulations at time $t=150$ s. This figure reveals a key feature of the CGP methodology. The pressure magnitude estimated by the IFE-CGP ($l=1$, and 2)  method is far from that computed with the full-fine scale resolution. Comparably, the pressure magnitudes obtained using the standard algorithm with the full fine and coarse resolutions are relatively close to each other. Notwithstanding this, the pressure gradient magnitude of the full fine-scale and the IFE-CGP ($l=1$, and 2) computations are indistinguishable for most of the domain (although with a phase lag in the case of one level coarsening). As discussed earlier, in contrast with the implicit pressure quantity, the pressure gradient is the relevant variable in the incompressible Navier-Stokes equations [27]. Hence, it does not matter if the IFE-CGP strategy does not retain absolute pressure values close to the full fine results. Importantly, it calculates the pressure gradients with a high level of accuracy and significantly better than those that are solely computed on a coarse grid.

\begin{figure*}
\centering
\includegraphics[width=165 mm]{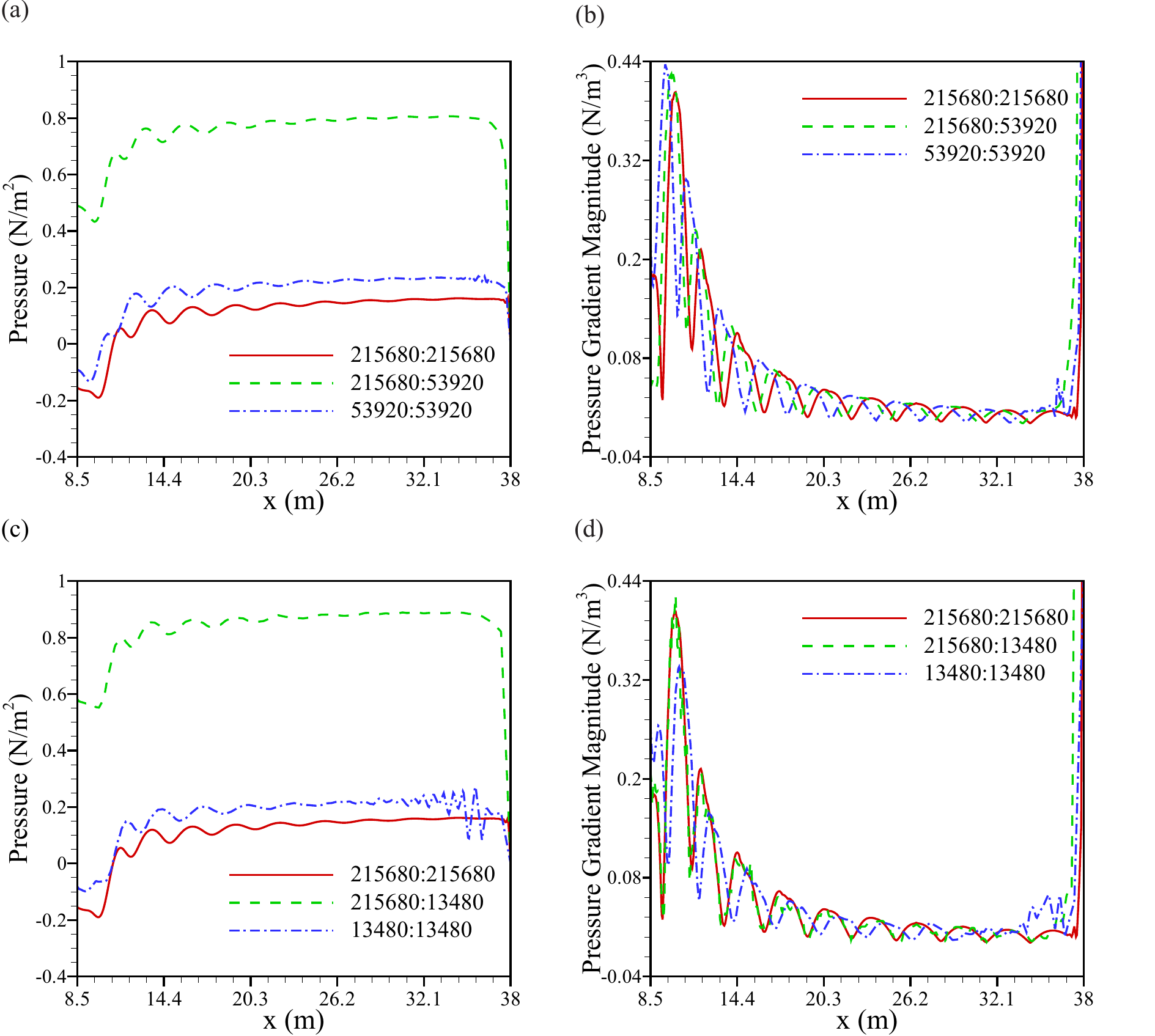}
\caption{The IFE-CGP methodology's effect on the integrity of \textbf{a} The pressure variable and \textbf{b} Its gradient; The 215680:215680 label indicates the standard algorithm with full fine scale computations, The 215680:53920 and 215680:13480 labels show, respectively, the IFE-CGP outputs with one and two levels of coarsening, and 53920:53920 and 13480:13480 labels illustrate the standard algorithm with full coarse scale computations.}
\label{fig:14}
\end{figure*}

So far we have emphasized the fact that spurious oscillations of the velocity and vorticity fields are removed by IFE-CGP. Now by looking at Fig. 14, we see that this is also the case for the pressure fields. Although both the IFE-CGP  ($l=1$, and 2) and full coarse (53920:53920 and 13480:13480) simulations solve the pressure Poisson equation on the same coarse mesh, the artificial oscillations at the end of domain are removed only in the case of the IFE-CGP outputs. Thus, for the same number of degrees of freedom, when the pressure Poisson equation is fed with a smoother intermediate velocity field, the outcome pressure filed is also smoother. Lastly, there is a noticeable difference between the pressure gradient magnitude of IFE-CGP and full fine scale computations near $x=38$ m. This difference comes from homogenous artificial Dirichlet boundary condition for the pressure ($p=0$). As can be seen from Fig. 14a and Fig. 14c, the pressure of IFE-CGP falls sharply near $x=38$ m to satisfy this boundary condition. It is conjectured that this issue would be eliminated by switching from non-incremental pressure projection methods to incremental ones.

\subsection{CGP as partial mesh refinement}
\label{sec:5}
Let's consider a condition that the standard numerical simulation diverged for the $N:N$ case due to a relative high Reynolds number or too coarse a mesh. Or the results obtained with a $N:N$ grid resolution are not sufficiently resolved and a fluid field with more detailed information is needed. The standard approach to resolving these common issues in projection methods is to refine both the advection-diffusion and the Poisson grids. In contrast with this approach, the CGP strategy suggests refining the advection-diffusion grid, without changing the resolution of the Poisson mesh. To be more precise from a terminology point of view, CGP does not propose a new mesh refinement method; however, it guides users to implement available mesh refinement techniques for the grids associated with the nonlinear equations. Here, we describe the concept by showing two simple examples.

Consider the simulation of the flow over a backward-facing step. Let's assume that one is interested in the flow information at $Re=800$; however, due to wall clock time or computational resource limitations, he is not able to run a simulation with the required pure fine 102400:102400 grid resolution. On the other hand, because a coarse 1600:1600 resolution is not high enough, the simulation diverges after 96 time steps, as depicted in Fig. 9a. The IFE-CGP framework with an intermediate resolution of 102400:1600 provides a converged solution as shown in Fig. 9b. The relative velocity error norms with reference to the full fine simulation are of order $10^{-5}$. Furthermore, the normalized reattachment length can be estimated around 14. Although this estimation is slightly different from that obtained by the standard computations, it is captured 30 times faster. Note that these results are achieved by only refining the advection-diffusion equation solver mesh. In this case, because the simulation on the coarse grid diverges, there is no real number for $C_c$; however, if a virtual $C_c$ considered, $h_f/h_{cgp}=29.534$.

Let's reconsider the flow over a circular cylinder computations described in Sect. 3.3. Coming back to the lift coefficient graphs, depicted in Fig. 13, and with an emphasis on the IFE-CGP ($l=1$) outputs, another interpretation of these results is discussed here. Let's assume an exact measurement of the lift coefficient is needed for a specific engineering purpose. Using standard methods, this can be accomplished using either 215680:215680 or 53920:53920 grid resolutions. An implementation with the finer grid produces a more precise answer. It could be a user’s incentive to locally/globally refine the full coarse mesh. Obviously, this mesh refinement ends in an increase in CPU time for the processing part of the simulation. In this case, our numerical experiments show that the increase is equal to 339271.6 s (over 94 hr). As mentioned earlier, having a coarse mesh only degrades the level of the viscous lift accuracy. In fact, instead of refining the grids associated with both the nonlinear and linear equations, a mesh refinement of the nonlinear part is enough alone. Hence, in order to increase the precision of the lift force, one can refine the advection-diffusion grid and keep the resolution of the Poisson mesh unchanged. In this case, the IFE-CGP grid refinement cost factor is $h_{cgp}=3.638$, whereas this factor for the regular mesh refinement is $h_f=11.088$, illustrating a considerable saving of computational resources.

As a last point, obviously the types of two dimensional flow simulations described here are not challenging computation problems today. These problems are merely used as examples to explain one of the features of the IFE-CGP algorithm. Practical applications of the IFE-CGP mechanism as a mesh refinement tool are expected to be useful for three dimensional flow simulations on parallel machines.

\subsection{Boundary conditions and data structure effects on CGP efficiency}
\label{sec:6}
Figures 15a--15d compare the CPU times consumed by various components of the processing segment for four test cases using different boundary conditions. The Taylor-Green vortex, flow over a backward-facing step, and flow past a cylinder are modeled by the IFE-CGP approach, while the double shear layer problem is simulated by the CGPRK3 technique [16]. Regarding the IFE-CGP strategy, by the coarsening level increment, the Poisson equation price lessens dramatically so that its portion becomes less than 5\% after just one level of coarsening. For that reason, a considerable speedup cannot be achieved after $l=2$. A similar trend occurs in application of the CGPRK3 method with the difference that the major reduction in the Poisson solver cost is obtained at $l=3$. The significant difference between the IFE-CGP and CGPRK3 algorithms is associated with the computational expense of the data transfer between the advection-diffusion and the Poisson grids. According to Fig. 15d, the CGPRK3 method allocates more CPU resources to mapping the data for each subsequent level of coarsening, and finally the mapping process costs overcome the Poisson equation costs at $l=3$. In contrast, this charge never exceeds 0.004\% of the total algorithm computational cost using the IFE-CGP strategy.

\begin{figure*}
\centering
 \includegraphics[width=175 mm]{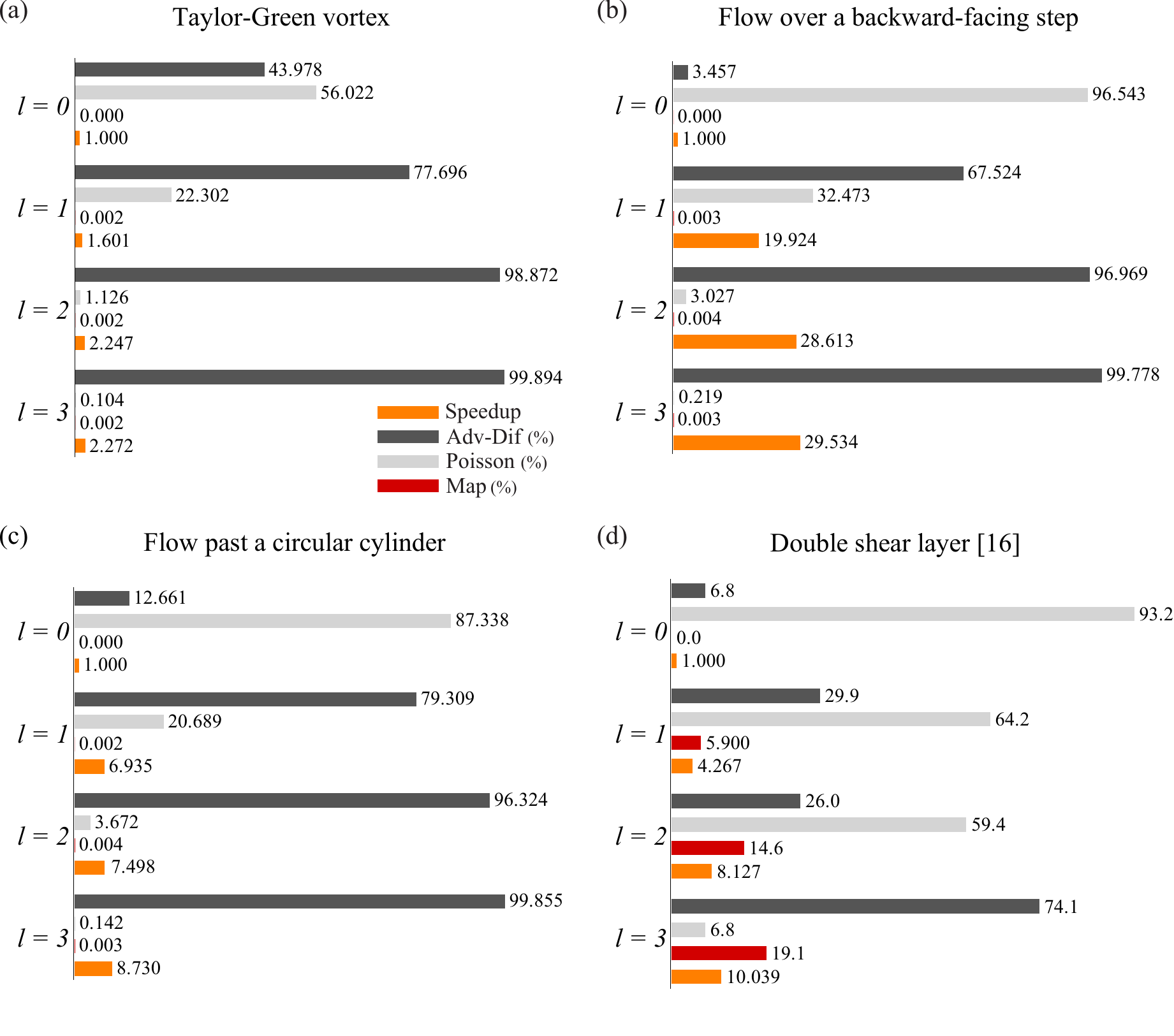}
\caption{The influence of boundary condition types and mapping algorithms on the achieved speedup by the CGP technique for \textbf{a} the Taylor-Green vortex problem with velocity Dirichlet boundary conditions,  \textbf{b} the flow over a backward-facing step with stress free conditions,  \textbf{c} the flow past a cylinder with open boundary conditions, and  \textbf{d} the double shear layer problem [16] with periodic boundary conditions.}
\label{fig:15}
\end{figure*}

\begin{figure}
\end{figure}

Concerning influences of the boundary condition, the maximum speedup is achieved when the outflow velocity in the backward-facing step problem is treated as a stress-free condition, and the lowest acceleration of the computations is observed when velocity Dirichlet boundary conditions are enforced for the Taylor-Green vortex simulation. The speedups when using periodic boundary conditions are in between these two extremes. This difference is expected because solving a Poisson equation with spurious homogeneous Dirichlet boundary conditions requires more computational effort in comparison with pressure Neumann conditions [56]. In terms of the accuracy level, the CGP class of methods retains the velocity field data close to that of full fine grid resolution simulations in the presence of less restrictive velocity boundary conditions, such as open and periodic ones. However, the CGP approach with pure velocity Dirichlet conditions results in a dampened velocity field, which has been also reported by Lentine et al. [15]. In the case of Euler's equations, where the viscous term is neglected and the flow is allowed to slip on the solid surfaces, the CGP method acquires much less damped flows. For the same reason, the damping phenomenon disappears when the CGP solver is run at high Reynolds numbers.

\section{Conclusions and future directions}
\label{sec:1}
The CGP method is a new multigrid technique applicable to pressure projection methods for solving the incompressible Navier-Stokes equations. In the CGP approach, the nonlinear momentum equation is evolved on a fine grid, and the linear pressure Poisson equation is solved on a corresponding coarsened grid. Mapping operators transfer the data between the grids. Hence, one can save a considerable amount of CPU time due to reducing the resolution of the pressure filed while maintaining excellent to reasonable accuracy, depending on the level of coarsening.

In this article, we proposed a semi-Implicit-time-integration Finite-element version of the CGP (IFE-CGP). The newly added semi-implicit time integration feature enabled CGP to run simulations with large time steps, and thus further accelerated the computations compared to the standard/previous CGP algorithms. The new data structure introduced resulted in nearly zero computational cost for the mapping procedures. Using the finite element discretization, CGP was adapted to be suitable for complex geometries and realistic boundary conditions. Moreover, the mapping functions were conveniently derived from the finite element shape functions.
 
In order to examine the efficiency of the IFE-CGP method, we solved three standard test cases: The Taylor-Green vortex in a non-trivial geometry, flow over a backward-facing step, and flow past a circular cylinder. The speedup factors ranged from 1.601 to 29.534. The minimum speedup belonged to the Taylor-Green vortex problem with velocity Dirichlet boundary conditions, while the maximum speedup was found for the flow over a backward-facing step with open boundary conditions. Generally, the outputs for one and two levels of the Poisson grid coarsening agreed well with those computed using full fine scale computations. For three levels of coarsening, however, only a reasonable level of accuracy was achieved.
 
The mesh refinement application of the CGP method was introduced for the first time in this work. Based on it, if for a given spatial resolution the numerical simulation diverged or the velocity outputs were not accurate enough, instead of refining both the advection-diffusion and the Poisson grids, the IFE-CGP mesh refinement suggests to only refine the advection-diffusion grid and keep the Poisson grid resolution unchanged. The application of the novel mesh refinement tool was shown in the cases of flow over a backward-facing step and flow past a cylinder. For the backward facing step flow, a three-level partial mesh refinement made a previously diverging computation numerically stable. For the flow past a cylinder, the error of the viscous lift force was reduced from 31.501\% to 7.191\% (with reference to the standard mesh refinement results) by the one-level partial mesh refinement technique.

We showed that the prolongation operator of IFE-CGP did not thicken the artificial layers that arose from the artificial Neumann boundary conditions. Additionally, we demonstrated that although CGP reduces the accuracy level of the pressure field, it conserves the accuracy of the pressure gradient, a key to the efficacy of the CGP method.

For future studies, the contribution of different incremental pressure projection schemes (such as standard [44], rotational [27], and vectorial forms [45]) to the CGP methodology will be analyzed. It is conjectured that because the above mentioned methods diminish the errors introduced by spurious Neumann conditions, applying a CGP technique to the incremental projection methods can lead to undamped flows in comparison with applying the CGP strategy to non-incremental pressure correction computations. Another objective of our future research is to perform a comparison between the CGP approach with one level of coarsening ($l=1$) and the standard finite element algorithm with Taylor-Hood mixed finite elements (\textbf{P}$_2$/\textbf{P}$_1$) (see e.g., Refs. [1, 6]). From a grid resolution point of view, for an assumed number of grid points of the velocity component, the Poisson solver utilizes a space with an equal pressure node numbers, discretized using either the IFE-CGP ($l=1$) method or Taylor-Hood elements. In this sense, a detailed investigation of the similarity/difference between these two concepts may introduce novel mapping functions for the IFE-CGP tool.


\begin{acknowledgements}
AK wishes to thank Dr. Michael Lentine, Dr. Peter Minev, Dr. Saad Ragab, and Dr. Omer San for helpful discussions.
\end{acknowledgements}



\end{document}